\documentclass[12pt]{article}
\usepackage[utf8]{inputenc}
\usepackage{geometry}
\usepackage{natbib}
\usepackage{amsmath}
\usepackage{amssymb}
\usepackage{bbm}
\usepackage{comment}
\usepackage{amsthm}
\usepackage{subcaption}
\usepackage[table,xcdraw]{xcolor}
\usepackage{graphicx}
\usepackage{booktabs}
\usepackage{array}
\usepackage{paralist}
\usepackage{verbatim}
\usepackage{multirow}
\usepackage{rotating}
\usepackage{mathtools}
\usepackage{doi}
\usepackage{fancyhdr}
\usepackage{bbm}
\usepackage{hyperref}
\usepackage[english]{babel}
\usepackage{amsthm}
\usepackage{tikz}
\theoremstyle{definition}
\usepackage{soul}

\theoremstyle{remark}

\usepackage{sectsty}
\allsectionsfont{\sffamily\mdseries\upshape}
\usepackage{algorithm}
\usepackage{algpseudocode}
\usepackage{authblk} 
\usepackage{fancyvrb}
\usepackage{lineno}
\usepackage{accents}

\providecommand{\keywords}[1]
{
  \small	
  \textbf{\textit{Keywords---}} #1
}
\title{\Large Extreme Value Inference with Measurement Error: Application to Earthquake Magnitudes}

\author{Wanchen Yue}
\author{Israel Martinez Hernandez}
\author{Jonathan A. Tawn}
\affil{\footnotesize Department of Mathematics and Statistics, Lancaster University LA1 4YF, United Kingdom.}
\author{Ross Towe}
\affil{\footnotesize Shell Information Technology International Limited, London SE1 7NA, United Kingdom}
\author{Zak Varty}
\affil{\footnotesize Department of Mathematics, Imperial College London SW7 2AZ, United Kingdom.}
\date{}
\begin{document}
\maketitle

\begin{abstract}
Extreme value analyses typically assume that observations are measured without error, yet many environmental measurements are subject to different levels of observational uncertainty depending on the context. Ignoring such measurement error can distort inference on the tail behaviour. In this paper, we focus on peaks-over-threshold models, where exceedances above a high threshold are typically modelled using the
generalised Pareto distribution (GPD). We develop a statistical framework for extreme value
inference when observations are noisy measurements of an underlying
true variable. Measurement error introduces a particular challenge in
threshold-based extreme value analysis, as latent exceedances may be
observed below the threshold, while latent non-exceedances may be observed
above it. To address this problem, we represent the latent distribution using a hybrid
body--tail specification, combining a flexible nonparametric body with
a GPD tail. We develop a computationally efficient partially censored likelihood inference approach for data of this structure. The framework allows the measurement error
scale and the GPD parameters to depend on covariates. Simulation
studies demonstrate biases on the tail
inference from ignoring measurement error and that our approach overcomes such impacts. The methodology is motivated by, and illustrated using, the catalogue of induced earthquake
magnitudes from the Groningen, Netherlands, one of the largest gas fields in the world.
\end{abstract}
\keywords{Earthquake Modelling, Extreme Value Theory, Fast Fourier Transform, Generalised Pareto Distribution (GPD), Groningen Gas Field, Measurement Error.}
\captionsetup{
    font=small
}

\section{Introduction}
\label{Sec: intro}
Extreme value methods are widely used to model rare and high-impact events across a range of scientific fields, particularly in environmental applications \citep{coles2001introduction, beirlant2006statistics, HandbookExtremes2026}. However, observations in these settings are often subject to measurement uncertainty arising from instrument precision, monitoring network geometry, and data-processing procedures. Consequently,
observations often represent noisy measurements of an underlying latent variable rather than of the quantity of direct interest. If such uncertainty is not explicitly accounted for, it will lead to biased estimates of the environmental hazard and mischaracterisation of the underlying true latent distribution. In addition, both the latent process and the accuracy of the
recorded measurements may vary with physical or environmental conditions and recording technology, so that the data need to be viewed as realisations of non-identically distributed variables. In such settings, variation in the observed data may reflect both genuine changes in the latent process and variability introduced by measurement uncertainty. Accounting for such covariate effects is therefore important for developing realistic statistical models and for separating signal from observational noise.

Measurement error has been extensively studied in statistics, with many methods developed to model uncertainty in observed data and to recover properties of an underlying latent variable. Early work focused on
linear measurement error models, where the effects of measurement error on regression and inference were analysed within the classical errors-in-variables framework \citep{fuller2009measurement}. In more general settings, \cite{carroll2006measurement} provide a comprehensive framework for
measurement error models, with an emphasis on bias reduction and
approximate consistency, including methods such as regression
calibration and simulation-based approaches. Meanwhile, nonparametric
methods based on deconvolution have been developed to recover the
distribution of a latent variable from noisy observations, often under
weaker distributional assumptions \citep{stefanski1990deconvolving, delaigle2008density}. These approaches have been applied in diverse areas, including
econometric, chemistry, and bioinformatics, where measurement uncertainty is a fundamental feature of the data. 

The role of measurement error in extreme value analysis has received comparatively limited attention. Almost all existing extreme value models are formulated under the assumption that observations are
measured without error, despite the fact that measurement uncertainty is often known to be present in practice for environmental data. A small number of studies have examined the role of measurement error in extreme value inference. For example, \cite{marra2018metastatistical} investigate the impact of measurement uncertainty on high-quantile estimation for rainfall data through simulation studies, whilst \cite{lin2023seeing} show that common sources of noise, such as rounding and discretisation, can bias inference for heavy-tailed distributions, and they propose logarithmic binning as a practical approach to mitigate these effects. Other related work has considered incomplete historical extreme event records, such as volcanic eruption data, where the probability of recording an event is modelled as a function of time and event size \citep{furlan2010extreme}. These studies highlight both the sensitivity of tail inference to observational uncertainty and the potential for tailored methods to address specific forms of error. Nevertheless, statistical methodology for extreme value analysis that explicitly accounts for measurement error does not exist for independent and identically distributed (iid) variables, nor for the non-identically distributed settings required for induced earthquake inference.

A common approach to modelling extremes is the peaks-over-threshold framework, in which observations exceeding a high threshold $u$ are modelled using the generalised Pareto distribution (GPD) \citep{davison1990models}. The corresponding distribution function is
\begin{equation}
F_{\mathrm{GPD}}(x;u,\sigma_u,\xi)
=
1-
\left(
1+\frac{\xi(x-u)}{\sigma_u}
\right)_{+}^{-1/\xi},
\qquad x>u,
\label{GPD_CDF}
\end{equation}
where $\sigma_u>0$ is the scale parameter at threshold $u$, $\xi \in \mathbb{R}$
is the shape parameter, and $(a)_+=\max(a,0)$. The shape parameter
determines the tail form: when $\xi<0$, the distribution has a finite
upper endpoint at $u-\sigma_u/\xi$; when $\xi=0$, the GPD reduces
to the exponential distribution; and when $\xi>0$, the distribution
has a heavy, unbounded upper tail. When observations are subject to measurement error, the extreme events identified from the observed data may be incorrectly classified: observations that exceed $u$ may correspond to latent values below $u$, while true latent variable exceedances may be recorded as observations below $u$. Moreover, because the threshold lies in the upper tail of the distribution, where the density of the latent variable is decreasing, a symmetric distribution of measurement error tends to produce more observations above the threshold
than there were for the latent variable, giving more apparent exceedances of $u$ than genuinely occur. Standard approaches that rely solely on observed exceedances therefore operate on a distorted and increased sample, in which genuine extremes may be missed and non-extreme observations may be incorrectly included. This misclassification then distorts the information contained in the tail data, which we will show lead to biased and falsely over-precise estimation of the tail parameters $(\sigma_u, \xi)$ and extreme quantiles, and thereby makes standard inference procedures \citep{davison1990models} for the analysis of extreme unreliable.

Inference for earthquake magnitudes is our motivation and provides a natural application of this measurement error framework. We focus on the
Groningen gas field in the Netherlands, one of the world's largest gas
fields, which has experienced induced seismicity associated with gas
extraction since the early 1990s with the largest recorded value of $3.6\,{M_L}$, in local magnitude units, in 2012 \citep{van2015induced}. In this setting, magnitudes are
inferred from seismic recordings and therefore are inherently subject to
measurement uncertainty. Since the local magnitude unit is
defined on a logarithmic amplitude scale, errors in magnitude are
naturally represented as additive errors on the $M_L$ scale. The errors for Groningen are anticipated by geoscientists to have a standard deviation of the order of
$0.1$--$0.3\,M_L$ \citep{dost2013august,bourne2020stress}. Despite this known measurement uncertainty, statistical
analyses commonly treat recorded magnitudes as the true value observations
\citep{bourne2018exponential, beirlant2019estimating, varty2021inference, murphy2025automated, yue2025integrating, Murphy2026spatio}. The precision of magnitude estimates in Groningen is not constant: the geophone network that is used to record earthquakes has expanded over time, leading to changes in sensor coverage and, consequently, in the accuracy of recorded magnitudes \citep{dost2017development}. This dependency of recording uncertainty on the monitoring network is not unique to seismic data. For example, in gridded rainfall datasets, uncertainty can depend on the density and availability of rain gauges used to construct grid-cell averages \citep{prakash2019new}. In earthquake applications, this changing accuracy motivates using information about the geophone network to model how measurement error varies between events. In addition, gas extraction influences the behaviour of the latent magnitude distribution \citep{bourne2020stress}. These features motivate a modelling approach in which the observed data are treated as the combination of an underlying latent process and a noise component, while allowing covariate-dependent structure in both the latent variable and the measurement error.

Based on advice from experts in seismic measurement, we develop a statistical framework that explicitly incorporates additive Gaussian measurement error into a peaks-over-threshold extreme value analysis \citep{rhoades2000effects, werner2008magnitude}, although the methodology can be easily extended to more
general measurement error structures and distributions. To address the misclassification issue about observations being above/below the threshold, inference cannot be based solely on threshold exceedances of the observed data. Instead, the observed variable is treated as a noisy measurement of an underlying latent variable, whose distribution is modelled using a hybrid body--tail specification, combining a flexible nonparametric representation for the body of the data with a GPD for the tail. Consequently, both the body and tail of the latent distribution contribute to the observed exceedances, so that a model for the full latent distribution is required rather than a tail-only specification. Inference is focused on the tail through a partially censored likelihood, which reduces sensitivity to the choice of the smoothing bandwidth in the nonparametric body component. Likelihood evaluation is carried out using the fast Fourier transform (FFT) based convolution, which enables efficient numerical computation of the integral defining the density of the observation. 

The remainder of the paper is organised as follows. Section~\ref{Sec:Modelsetup} introduces the statistical model, including the latent variable formulation, the hybrid body-tail distribution, and the covariate-dependent specifications for both the signal and error scales. The inference methodology is then described in Section~\ref{Sec:Inference}, covering the partially censored likelihood formulation, the FFT-based evaluation of the convolution likelihood and the regularisation used to stabilise estimation in the presence of correlated covariate effects. A simulation study is presented in Section~\ref{Sec:Simulationstudy} to assess the performance of the proposed approach, relative to fitting the GPD to observations and the idealised case when there is no measurement error. This study covers varying levels of signal to noise ratio and different covariate structures. The methodology is illustrated in Section~\ref{Sec: Application} using the earthquake magnitude data from the Groningen induced seismicity catalogue. Finally, Section~\ref{Sec:discussion} concludes with a discussion of the results and directions for future work.

\section{Model setup}
\label{Sec:Modelsetup}
\subsection{Observed and latent variable model}
\label{subsec:observedandlatent}

We consider a measurement error setting in which the variable of
interest is not directly observed. Instead, inference must be based on
noisy measurements of an underlying latent quantity. Motivated by the earthquake magnitude setting, described in Section~\ref{Sec: intro}, we use an additive Gaussian measurement error model,
which is natural for magnitudes recorded on a logarithmic scale. Let $X_i$ denote the unobserved true latent value, and let $Y_i$ denote a noisy observation of $X_i$ for observations $i=1,\dots,n$, and we use the notation $\boldsymbol{Y}=(Y_1, Y_2, \ldots, Y_n)$ and $\boldsymbol{X}=(X_1, X_2, \ldots, X_n)$ for samples of size $n$ for the respective variables. The primary objective is to make inference on the
distribution of the latent variables $\boldsymbol{X}$, with a particular focus on its marginal upper tail behaviour, based on the observed data $\boldsymbol{y}=(y_1,y_2,\dots,y_n)$ on $\boldsymbol{Y}$.

To allow for non-identically distributed observations, we consider univariate covariates $Z_{L,i}$ and $Z_{N,i}$, for $i = 1, \ldots, n$, which influence the latent distribution and the measurement error
distribution, respectively. Let
$
\boldsymbol{Z}_i = (Z_{L,i}, Z_{N,i}), 
~
\boldsymbol{Z} = (\boldsymbol{Z}_1,\ldots,\boldsymbol{Z}_n),
$
where $\boldsymbol{Z}_i$ denotes the covariate information associated with observation $i$, and $\boldsymbol{Z}$ denotes the full collection of covariates. The corresponding parameter vectors are
denoted by $\boldsymbol{\theta}_L$ and $\boldsymbol{\theta}_N$, and we write
$
\boldsymbol{\theta}
=
(\boldsymbol{\theta}_L,\boldsymbol{\theta}_N).
$
We assume that, conditional on these covariates, the pairs $(X_i, Y_i)$ are
independent over $i = 1, \ldots, n$. In addition, the measurement error $\varepsilon_i$ is
assumed to be independent over $i$ and of the latent variables $X_i$ for $i=1,\dots,n$, conditional on $\boldsymbol{Z}_i$. The observed variable $Y_i$ is linked to $X_i$ by $\varepsilon_i$ the measurement error for the $i$-th observation through the additive measurement error model
\begin{equation}
\begin{aligned}
X_i \mid Z_{L,i} &\sim f_{X_i}(\,\cdot\,; \boldsymbol{\theta}_L, Z_{L,i}),
\qquad
Y_i = X_i + \varepsilon_i,\\[3pt]
\text{where} \quad \varepsilon_i \mid Z_{N,i} &\sim \mathcal N\!\bigl(0, \sigma_{N,i}^2\bigr),
\qquad
\sigma_{N,i} = g_N(Z_{N,i}; \boldsymbol{\theta}_N),
\qquad i = 1, \ldots, n,
\end{aligned}
\label{eq:mainmodel}
\end{equation}
for a suitable link function $g_N(\cdot) \geq 0 $ and parameter vector
$\boldsymbol{\theta}_N$. Here, $f_{X_i}(\,\cdot\,; \boldsymbol{\theta}_L, Z_{L,i})$ denotes the latent
density for observation $i$, defined in Section~\ref{subsec:latentbodytail}, and the conditional distribution of the observed variable $Y_i \mid X_i=x$ has density function
\begin{equation}
\phi\bigl(y - x; \sigma_{N,i}\bigr), \quad \text{for} \quad y\in \mathbb{R}
\label{eq:cond_error}
\end{equation}
where $\phi(\cdot;\sigma_{N,i})$ denotes the Gaussian density with
mean zero and standard deviation $\sigma_{N,i}$. 

\subsection{Latent body--tail model and observed density} 
\label{subsec:latentbodytail}
In the measurement error setting of Section~\ref{subsec:observedandlatent}, the latent variables
$\boldsymbol{X}$ are not directly observed, and inference must instead
be based on the noisy observations $\boldsymbol{Y}$. Because
measurement error can move observations across a threshold $u$ relative
to the latent variable, the observed values near to $u$ may be influenced by both the body and tail of the latent distribution.
As a result, when modelling only observations above $u$, it is
generally insufficient to assume a model for $f_{X_i}$ above that threshold only, and a coherent model for the full latent density
$f_{X_i}$ is required, which we specify next.

Since our primary interest lies in upper-tail behaviour, extreme value
theory provides a natural framework for modelling the tail of
$f_{X_i}$ \citep{pickands1975statistical,coles2001introduction}. In particular, for a sufficiently high threshold $u$, the
conditional distribution of exceedances of the latent variable is
modelled using a GPD, so that
\begin{equation}
(X_i-u)\mid (X_i>u,Z_{L,i})
\sim \mathrm{GPD}(\sigma_{u,i},\xi),
\label{eq:dependentGPD}
\end{equation}
where $\sigma_{u,i}$ is the observation-specific scale parameter for
threshold choice $u$ depending on $Z_{L,i}$, and $\xi$ is taken to be common across observations.
Using the notation introduced in distribution~\eqref{GPD_CDF}, the corresponding
density is
\[
f_{\mathrm{GPD}}(x;u,\sigma_{u,i},\xi)
=
\frac{1}{\sigma_{u,i}}
\left(
1+\frac{\xi(x-u)}{\sigma_{u,i}}
\right)_{+}^{-\left(1/\xi+1\right)},
\qquad x>u.
\]
We allow the scale parameter to depend on covariates through
$
\sigma_{u,i}=g_L(Z_{L,i};\boldsymbol{\theta}_L),
$
for a suitable link function $g_L(\cdot) \geq 0$ and parameter vector
$\boldsymbol{\theta}_L$. The choice that $\xi$ is constant is
standard in many applied GPD analyses, reflecting both the limited
evidence typically available for systematic covariate effects in the
shape parameter and the practical difficulty of estimating even a
homogeneous shape parameter reliably \citep{healy2025inference} coupled with reducing the risk of parameter
identifiability issues \citep{davison2012statistical}.

The GPD specification in model~\eqref{eq:dependentGPD} describes the tail, but as mentioned above, the full latent distribution is needed for inference under measurement error. We therefore build on the marginal body--tail
modelling framework of \citet{coles1991modelling}. In that framework, the body of the distribution is modelled empirically or using a smooth kernel density estimate, and is joined to a GPD tail so that the resulting distribution function is continuous at the threshold. Here, density continuity is also desirable because earthquake magnitude is treated as a continuous quantity, and there is no geophysical reason to assign positive probability mass to any particular non-zero magnitude value. Therefore, a sharp transition between the body density and the GPD tail density may therefore make the convolved
observed density (see below in expression~\eqref{eq:densityy}) sensitive to the density discontinuity at the threshold. For these reasons, we modify the standard body--tail construction by introducing a
transition region in which the latent density changes smoothly and continuously from the
nonparametric body component to the GPD tail component. To do this, we introduce a transition width $\delta \ge 0$. The threshold $u$ marks the point from which the covariate-dependent GPD approximation in \eqref{eq:dependentGPD} is considered plausible, while
$u+\delta$ is the threshold above which we assume the GPD tail model is
used fully. In practice, the threshold at which the latent distribution begins to
follow a GPD is unknown since the latent values are not
directly observed, the interval $(u,u+\delta)$ therefore provides a
transition region that reflects uncertainty about where the GPD tail
should begin to dominate. 

For the non-extreme component of the $i$-th latent distribution, we use a
flexible nonparametric density estimate $f_{\mathrm{NP},i}$,
constructed using a kernel density estimator (KDE). Let $K(\cdot)\geq 0$ denote a kernel function, $b>0$ denote
the bandwidth that controls the smoothness of the estimated density \citep{silverman2018density}, and let
$\mathcal{I}_i$ denote the set of values
used to construct the nonparametric density associated with latent
variable $i$. If the latent variable was observed without error, i.e., with values $(\tilde{x}_1,\dots, \tilde{x}_n)$, then
\begin{equation}
f_{\mathrm{NP},i}(x)
=
\frac{1}{|\mathcal{I}_i| b}
\sum_{j \in \mathcal{I}_i}
K\left(\frac{x-\tilde{x}_j}{b}\right),
\qquad x\in\mathbb{R}.
\label{eq:np_formula}
\end{equation}
Because the latent variables are not directly observed, we take $\tilde{x}_i=y_i$ for $i =1,\dots, n$, a decision which has minimal impact on inference relative to the ideal case of setting $\tilde{x}_i=x_i$, see Table~\ref{tab:rmse_efficiency}.

The body--tail hybrid construction uses this nonparametric density for
the body part of the latent distribution and the unconditional GPD density for the
upper tail, i.e., $x \leq u$ and $x \geq u+\delta$ respectively. For $x\in(u,u+\delta)$, we adopt a weighted transition
between the two, with mixing
probability $w_\delta(x)=(u+\delta-x)/\delta$, for $\delta>0$. We write the latent hybrid density for observation
$i$ as $f_{X_i}(x; \boldsymbol{\theta}_L)$, which is defined as being proportional to (up to the normalisation constant)
\begin{equation}
 f_{X_i}(x;\boldsymbol{\theta}_L) \propto
\begin{cases}
f_{\mathrm{NP},i}(x), & x \le u, \\[4pt]
w_\delta(x)\,f_{\mathrm{NP},i}(x)
+ \{1-w_\delta(x)\}\lambda_{u+\delta,i}
f_{\mathrm{GPD}}(x;u,\sigma_{u,i},\xi),
& u < x < u+\delta, \\[4pt]
\lambda_{u+\delta,i}
f_{\mathrm{GPD}}(x;u,\sigma_{u,i},\xi), & x \ge u+\delta,
\end{cases}
\label{eq:latentmodel}
\end{equation}
where the tail scaling constant $\lambda_{u+\delta,i}$ is defined by
\[
\lambda_{u+\delta,i}
=
\frac{1-F_{\mathrm{NP},i}(u+\delta)}
     {1-F_{\mathrm{GPD}}(u+\delta;u,\xi,\sigma_{u,i})},
\]
with
\[
F_{\mathrm{NP},i}(x)
=
\int_{-\infty}^{x}
f_{\mathrm{NP},i}(t)\,\mathrm{d}t,
\qquad x\in\mathbb{R}.
\]
This choice of $\lambda_{u+\delta,i}$ ensures self-consistency of the tail probabilities from the two
components of the model. When $\delta=0$, $
\lambda_{u+\delta,i}=\lambda_{u,i}
=
1-F_{\mathrm{NP},i}(u),
$
which corresponds to the standard body--tail construction of
\citet{coles1991modelling}, where the tail component begins directly at
the threshold $u$ without an intermediate transition region. Construction~\eqref{eq:latentmodel} preserves the GPD tail behaviour above $u+\delta$ while providing a coherent
continuous density model for the latent variable, which it is anticipated should provide better inference under measurement error.

Given this latent density~\eqref{eq:latentmodel}, the marginal distribution of the observed
variable $Y_i$ is obtained by integrating out the latent variable
$X_i$. Combining the hybrid density $f_{X_i}$ with the conditional
measurement error model \eqref{eq:cond_error}, the observed density is
given by the convolution
\begin{equation}
f_{Y_i}(y;\boldsymbol{\theta},\boldsymbol{Z}_i)
=
\int_{-\infty}^{\infty}
f_{X_i}(x;\boldsymbol{\theta}_L,Z_{L,i})\,
\phi\{y-x;\sigma_{N,i}\}\,\mathrm{d}x,
\quad y \in \mathbb{R},
\quad i=1,\ldots,n .
\label{eq:densityy}
\end{equation}
We term the model formulation~\eqref{eq:densityy} the GPD-ME model. Conditional on $\boldsymbol{Z}$, the observations are
independent, so the full observed-data density is
\begin{equation}
f_{\boldsymbol{Y}}(\boldsymbol{y};\boldsymbol{\theta},\boldsymbol{Z})
=
\prod_{i=1}^{n}
f_{Y_i}(y_i;\boldsymbol{\theta},\boldsymbol{Z}_i).
\label{eq:prodydensity}
\end{equation}
This product form provides the basis for likelihood-based inference.
However, direct evaluation of expression \eqref{eq:prodydensity} can be computationally
demanding when both the latent density $f_{X_i}$ and the measurement
error density $\phi(\cdot;\sigma_{N,i})$ vary across observations through covariates. The results can be sensitive to the choice of the nonparametric body bandwidth as the bandwidth can be confounded with measurement error. Sensitivity can also arise from model misspecification, for example if the measurement error variance changes with the
value of $X$ for non-extreme values of $X$. We therefore propose a range of approaches to overcome these issues in Section~\ref{Sec:Inference}.

\section{Inference for the GPD-ME model}
\label{Sec:Inference}
\subsection{Partially censored likelihood}
\label{subsec:pcens_likelihood}

Although likelihood~\eqref{eq:prodydensity} defines a full joint density for the observed
variable for our convolution model, our primary interest is inference on the upper tail of the
latent distribution. Given our reservations about using likelihood~\eqref{eq:prodydensity}, as expressed in Section~\ref{subsec:latentbodytail}, we focus only on the extreme observations by using a partially censored likelihood that
treats observations below the tail threshold $u+\delta$ as censored \citep{smith1997markov}. This
retains information about the probability of observations falling below
the threshold, while avoiding the need to model their actual values. 

In the censored likelihood, observed values above the threshold $u+\delta$ contribute through the convolved
density, whereas observations below the threshold contribute only
through the probability of lying below $u+\delta$. The partially censored likelihood contribution of observation
$i$ is
\[
L_i(\boldsymbol{\theta})
=
\begin{cases}
F_{Y_i}(u+\delta;\boldsymbol{\theta},\boldsymbol{Z}_i),
& y_i \le u+\delta, \\[6pt]
f_{Y_i}(y_i;\boldsymbol{\theta},\boldsymbol{Z}_i),
& y_i > u+\delta .
\end{cases}
\]
The corresponding overall likelihood is therefore
\begin{equation}
L(\boldsymbol{\theta})
=
\prod_{i:y_i \le u+\delta}
F_{Y_i}(u+\delta;\boldsymbol{\theta},\boldsymbol{Z}_i)
\prod_{i:y_i > u+\delta}
f_{Y_i}(y_i;\boldsymbol{\theta},\boldsymbol{Z}_i).
\label{eq:Censoredlikeli}
\end{equation}

\subsection{Convolution-based likelihood evaluation}
\label{subsec:fft_likelihood}

The density of the observations defined in likelihood~\eqref{eq:Censoredlikeli}, provides the
basis for likelihood-based inference under the proposed measurement error
model. This places the problem within a deconvolution framework, since
inference on the latent distribution must be performed indirectly through
noisy observations rather than from a direct sample of the variable of
interest
\citep{stefanski1990deconvolving,delaigle2008density}. While the evaluation of the convolutions for likelihood~\eqref{eq:Censoredlikeli} is conceptually straightforward, direct
numerical evaluation for every observation above $u+\delta$ and every
parameter update would generally be computationally expensive. We
therefore evaluate the convolution using the FFT \citep{nussbaumer1981fast}. This avoids repeated direct numerical
integration and provides an efficient route for evaluating the observed
density and distribution function required in the likelihood. After discretisation on an equally
spaced grid, the convolution theorem allows the convolution integral to
be approximated by multiplication in the Fourier domain
\citep{silverman}.
For observation $i$, the convolution is approximated on a common equally spaced grid
$
x_1,\ldots,x_J,
$
with spacing $\Delta x$. The grid limits $x_1$ and $x_J$, the
spacing $\Delta x$, and hence the number of grid points $J$, are
chosen by the user to provide a sufficiently accurate numerical
approximation over the range where the latent and observed densities
have non-negligible mass. For the $j-$th grid, we define the discretised latent
density and measurement error density by
\[
f_{X_i,j}
=
f_{X_i}(x_j;\boldsymbol{\theta}_L,Z_{L,i}),
\qquad
\phi_{i,j}
=
\phi(x_j;\sigma_{N,i}),
\qquad j=1,\ldots,J .
\]
Let $\boldsymbol{f}_{X_i}=(f_{X_i,1},\ldots,f_{X_i,J})$ and
$\boldsymbol{\phi}_{i}=(\phi_{i,1},\ldots,\phi_{i,J})$ denote the corresponding
vectors of density values on the grid.

After evaluating both densities on the grid, the convolution~\eqref{eq:densityy} is approximated by a discrete convolution of the two
vectors of grid values. For two sequences $\boldsymbol{a}=(a_1,\ldots,a_J)$ and
$\boldsymbol{b}=(b_1,\ldots,b_J)$, we write $\boldsymbol{a}*\boldsymbol{b}$ for their discrete convolution, $\mathcal{F}$ denoting the discrete Fourier transform and
$\mathcal{F}^{-1}$ denoting its inverse. The convolution theorem states that the discrete Fourier transform of a convolution is equal to the pointwise product of the individual discrete Fourier
transforms \citep{silverman}, namely
$
\mathcal{F}\{\boldsymbol{a}*\boldsymbol{b}\}
=
\mathcal{F}\{\boldsymbol{a}\}\,\mathcal{F}\{\boldsymbol{b}\}.
$
Therefore, the discretised
observed density for $Y_i$ can be computed by applying the inverse
discrete Fourier transform to the pointwise product of the transformed
latent density and transformed measurement error density, i.e.,
$
\boldsymbol{f}_{Y_i}
\approx
\Delta x\,
\mathcal{F}^{-1}
\left[
\mathcal{F}(\boldsymbol{f}_{X_i})
\cdot
\mathcal{F}(\boldsymbol{\phi}_{i})
\right],
$
where $\cdot$ denotes pointwise multiplication of the two transformed
vectors.

\subsection{Covariate discretisation for computational issues}
\label{subsec:covariate_discretisation}

The partially censored likelihood in Section~\ref{subsec:pcens_likelihood}
is written in terms of observation-specific densities $f_{Y_i}$ and
distribution functions $F_{Y_i}$. When the latent distribution and the
measurement error distribution vary continuously across observations
through covariates, evaluating these quantities separately for every
observation can remain computationally expensive, even with the FFT-based
convolution. We therefore use a discretised approximation to account for the
covariate effects.

Let $\mathcal{B}_{L,1},\ldots,\mathcal{B}_{L,G_L}$ denote a partition
of the range of the latent covariate $Z_L$, and let 
$\mathcal{B}_{N,1},\ldots,\mathcal{B}_{N,G_N}$ denote a partition of
the range of the measurement error covariate $Z_N$. For observation
$i$, let $I_L(i)$ and $I_N(i)$ denote the indices of the bins
containing $Z_{L,i}$ and $Z_{N,i}$, respectively. For $r=1,\ldots,G_L$ and $s=1,\ldots,G_N$, let
$Z_{L,r}^{\mathrm{rep}}$ and $Z_{N,s}^{\mathrm{rep}}$ denote
representative values of $Z_L$ in $\mathcal{B}_{L,r}$ and $Z_N$ in
$\mathcal{B}_{N,s}$, respectively.

For each latent covariate bin $\mathcal{B}_{L,r}$, the nonparametric body component used in $f_{X_i}$ is estimated separately using observations in that bin. Thus, $f_{X_i}$ may differ between $Z_L$-bins through both the KDE body component and the GPD scale
evaluated at $Z_{L,r}^{\mathrm{rep}}$. The observation-specific density
$f_{Y_i}$ is approximated by
\[
\widetilde f_{Y_i}(y;\boldsymbol{\theta})
=
\int_{-\infty}^{\infty}
f_{X_i}\!\left(
x;\boldsymbol{\theta}_L,Z_{L,I_L(i)}^{\mathrm{rep}}
\right)
\phi\!\left\{
y-x;
g_N\!\left(Z_{N,I_N(i)}^{\mathrm{rep}};\boldsymbol{\theta}_N\right)
\right\}
\,\mathrm{d}x,
\]
with the corresponding distribution function denoted by
$\widetilde F_{Y_i}(y;\boldsymbol{\theta})$. Both
$\widetilde f_{Y_i}$ and $\widetilde F_{Y_i}$ are evaluated using
the FFT procedure described in Section~\ref{subsec:fft_likelihood} with the number of FFT evaluation is reduced to $G_L G_N$ types for each $i$. Under this discretised approximation, the partially censored likelihood
becomes
\begin{equation}
L_{\mathrm{disc}}(\boldsymbol{\theta})
=
\prod_{i:y_i \le u+\delta}
\widetilde F_{Y_i}(u+\delta;\boldsymbol{\theta})
\prod_{i:y_i > u+\delta}
\widetilde f_{Y_i}(y_i;\boldsymbol{\theta}),
\label{eq:disc_censored_likelihood}
\end{equation}
which preserves covariate-dependent structure in both the
latent distribution and the measurement error distribution, while
avoiding fully observation-specific convolution at each likelihood
evaluation.

The numbers of bins, $G_L$ and $G_N$, and the bin boundaries are
numerical choices. Increasing the number of bins gives a closer
approximation to the fully observation-specific likelihood, but also
increases the number of convolutions that must be evaluated. In
practice, these choices should balance numerical accuracy and
computational efficiency. Sensitivity to the discretisation can be
assessed by repeating the analysis with alternative bin numbers or bin
boundaries; the choices used in the Groningen application are reported
in Section~\ref{Sec: Application}.

\subsection{Penalisation for covariate identifiability}
\label{subsec:penalisation}

The covariate-dependent formulation allows both the latent distribution
and the measurement error distribution to vary across observations. This
flexibility is important in applications where the underlying process
and the observation process are both non-identically distributed.
However, when the covariates affecting these two components are strongly
associated, or exhibit similar patterns over the observation period,
their effects can be difficult to distinguish from the likelihood alone.
In particular, variation in the observed upper tail may be attributed to
changes in the latent tail distribution, changes in the measurement
error distribution, or a combination of both. This can lead to weak
identifiability when both components are allowed to vary with
covariates.

To address this issue, we propose penalising the measurement error
component. The purpose of the penalty is to restrict the measurement
error scale to a scientifically plausible range and to reduce
confounding between the latent tail scale and the measurement error
scale. Such information is often available from the measurement context:
e.g., instrument precision, or expert
knowledge may provide guidance on plausible values of the measurement
error scale. The penalty is therefore applied to the parameters
$\boldsymbol{\theta}_N$ governing the measurement error scale through
the link function $g_N$ in expression~\eqref{eq:mainmodel}. Therefore, we maximise a penalised version of the discretised partially censored likelihood. Let
$
\ell_{\mathrm{disc}}(\boldsymbol{\theta})
=
\log L_{\mathrm{disc}}(\boldsymbol{\theta}),
$
where $L_{\mathrm{disc}}(\boldsymbol{\theta})$ is defined in
\eqref{eq:disc_censored_likelihood}. We consider the penalised objective function
\begin{equation}
\ell_p(\boldsymbol{\theta})
=
\ell_{\mathrm{disc}}(\boldsymbol{\theta})
-
P_N(\boldsymbol{\theta}_N),
\label{eq:penalised_likelihood}
\end{equation}
where $P_N(\boldsymbol{\theta}_N)$ is a penalty on the measurement
error component. The penalised estimator is then
$
\widehat{\boldsymbol{\theta}}
=
\arg\max_{\boldsymbol{\theta}}
\ell_p(\boldsymbol{\theta}).
$ The specific form of $P_N$ depends on the available scientific
information and on the application. In Section~\ref{Sec: Application},
we specify $g_N$ and the $P_{N}$ used for the Groningen
earthquake catalogue.

\section{Simulation Study}
\label{Sec:Simulationstudy}

\subsection{Simulation Design}
\label{subsec:simulation}

We conduct a simulation study to assess the performance of the proposed GPD-ME approach (i.e., model and inference) under three controlled settings (scenarios) motivated by the Groningen earthquake magnitude application in Section~\ref{Sec: Application}. The parameter
values are chosen to be comparable with those estimated for the Groningen
magnitude data, so that the simulated datasets have realistic tail
behaviour and measurement error levels. Scenario~1 considers a
baseline iid setting in which both the GPD scale and measurement error scale
are constant. Scenario~2 allows both scales to depend on independently
generated covariates $(Z_L, Z_N)$. In the third scenario, reported in the supplementary material, the two covariates are sampled from the joint distribution of the dependent covariates $(Z_L, Z_N)$ in Groningen analysis, see Figure~\ref{fig:groningen_covariates}. No
penalisation is used in scenarios~1 and~2, whereas the correlated-covariate setting in scenario~3 considers the regularisation scheme described in Section~\ref{subsec:penalisation}. The aim is to examine the quality of inferences we can achieve using the GPD-ME when the observations are contaminated by additive Gaussian measurement error. We compare the proposed GPD-ME method, fitted to the noisy observations $Y_i$ using the convolution-based likelihood, with a na\"ive GPD fit that models the noisy observations above $u+\delta$ as exact threshold exceedances. In scenario~1, we also include an oracle GPD fit to the latent variable $X_i$ above $u+\delta$. The oracle fit is only available in simulation as there the true $X_i$ values are known, but provides a useful
reference for separating the effect of measurement error from ordinary sampling variability.

Each simulated dataset contains $n=2000$ latent variables $(X_1,\ldots,X_n)$ generated from the hybrid body--tail distribution in expression~\eqref{eq:latentmodel}, using the lower threshold $u=0.76$ and upper threshold $u+\delta=1.16$ as in the Groningen application and we generate $200$ independent datasets. The nonparametric body component $f_{\mathrm{NP}}$ is fixed across all simulation scenarios. To construct
this component, we first generate samples from a normal distribution with mean $0.4$ and standard deviation $0.35$. A kernel density estimate (KDE) is then computed on a grid with spacing $0.01$. The
GPD tail and transition region are then defined as in expression~\eqref{eq:latentmodel}, and the resulting hybrid density is normalised to integrate to one. This choice gives a smooth body distribution with most mass at small magnitudes, broadly matching the non-extreme part of the Groningen magnitude distribution.

In the main simulation experiments, the nonparametric body component
below $u+\delta$ is treated as known and fixed, so that the study focuses on
inference for the GPD tail and measurement error components. To assess
the sensitivity to this modelling choice, we also consider a
more realistic variant in scenario~1 in which the body KDE is estimated
from the noisy observations $\boldsymbol{Y}$, rather than
from the latent variables $\boldsymbol{X}$. Due to our use of a censored likelihood, we expect the choice of body KDE to have a limited but non-negligible effect
on inference. The corresponding results are
reported as a sensitivity analysis in scenario~1 and are omitted from
the later scenarios to avoid repetition.

Observed noisy data $(Y_1, \ldots, Y_n)$ are obtained by adding Gaussian measurement error independently to
the latent variables. The simulations
are designed to assess inference under different signal-to-noise
settings, sensitivity to the GPD parameters $\xi$ and $\sigma_{u+\delta}$, as well as under covariate variation in the latent GPD scale $\sigma_{u+\delta,i}$
and measurement error scale $\sigma_{N,i}$. The scale functions and covariates used in these two components are
specified separately for each simulation scenario.

\subsection{Scenario 1: Constant signal and noise scales}
\label{subsec: scenario1}
The first simulation scenario considers the baseline iid setting in which both the latent GPD scale and the measurement error scale are constant. We consider three noise levels, $\sigma_N\in\{0.1,0.2,0.4\}$, to assess the impact of the signal-to-noise ratio on tail inference. The range is chosen to be broadly consistent with the levels of magnitude error scale considered plausible for the Groningen catalogue. The true tail parameters are $\xi=-0.15$ and $\sigma_{u+\delta}=0.6$. These values are also chosen to be close to those for the Groningen magnitude application. To address the broader impact of the GPD shape parameter, additional results for $\xi=0.15$ and for a near-exponential case with $\xi=-0.01$ are reported in the supplementary material.

Figure~\ref{fig:sim_case1_tail_params} shows the sampling distributions of the fitted shape and scale parameter estimates across the replicates for the GPD-ME, na\"ive and oracle methods for each of the three noise levels. When the noise level is small, $\sigma_N=0.1$, the three methods give similar estimates for each of $\xi$ and $\sigma_{u+\delta}$. The differences between the inference performance becomes clearer as the signal-to-noise level decreases. The na\"ive method increasingly underestimates $\xi$ and overestimates $\sigma_{u+\delta}$, with the stronger distortion occurring as $\sigma_N$ increases. In contrast, the GPD-ME estimators for both scale and shape parameters remain closer to the oracle fit, although its sampling variability increases as the noise level becomes larger. Numerical summaries of bias, variance and RMSE for these parameter estimates are reported in the supplementary material.

\begin{figure}[H]
    \centering
    \includegraphics[width=0.95\linewidth]{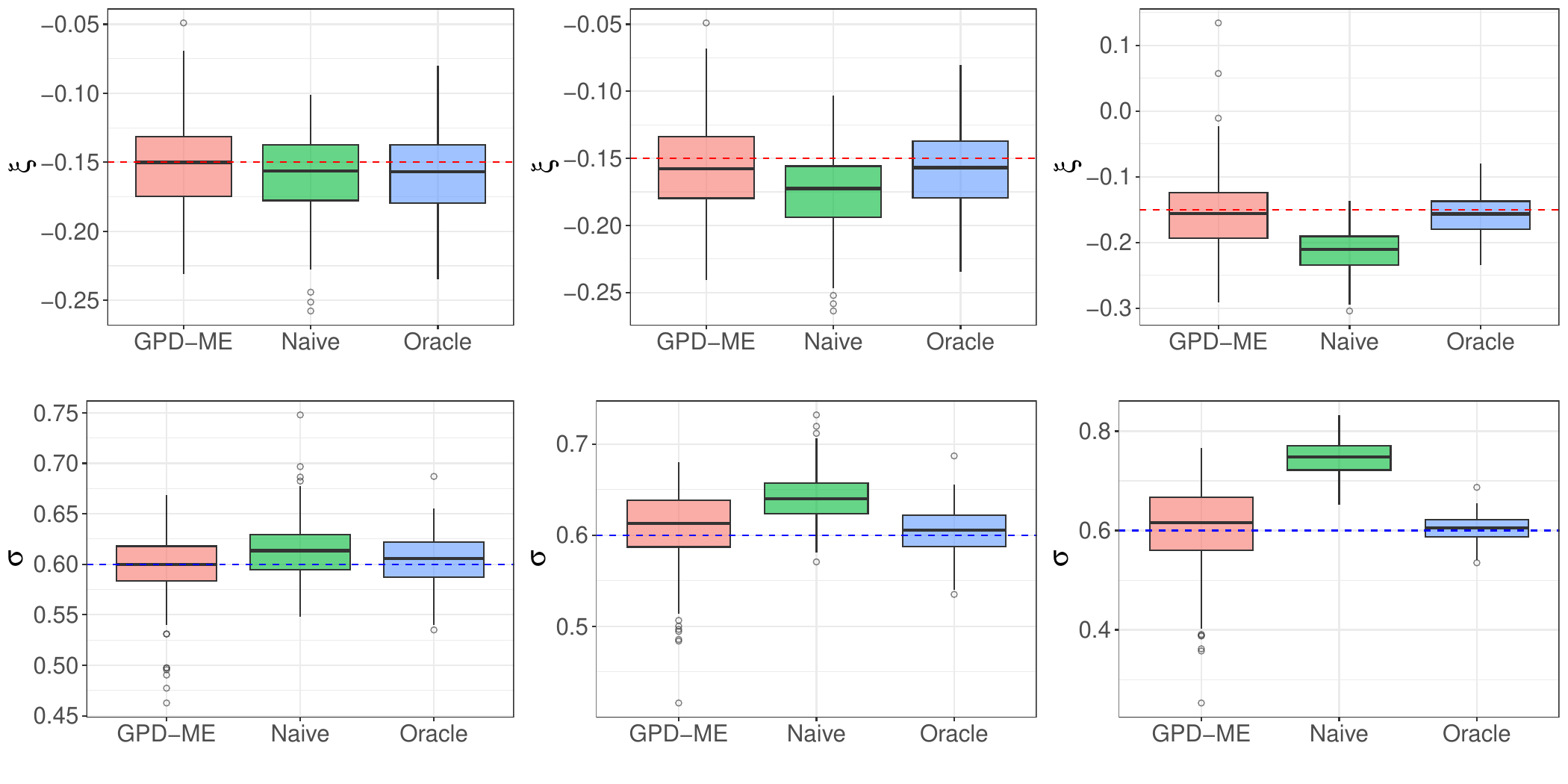}
\caption{Sampling distributions of the fitted GPD shape parameter $\xi$ (top row) and
scale parameter $\sigma_{u+\delta}$ (bottom row) in  scenario 1 for different signal-to-noise settings,
$\sigma_N\in\{0.1,0.2,0.4\}$ (left to right columns). Results are shown for the GPD-ME (red), na\"ive (green) and oracle (blue) methods. Dashed horizontal lines indicate the
true parameter values.} 
\label{fig:sim_case1_tail_params} 
\end{figure}

\begin{figure}[H]
\centering
\includegraphics[width=\textwidth]{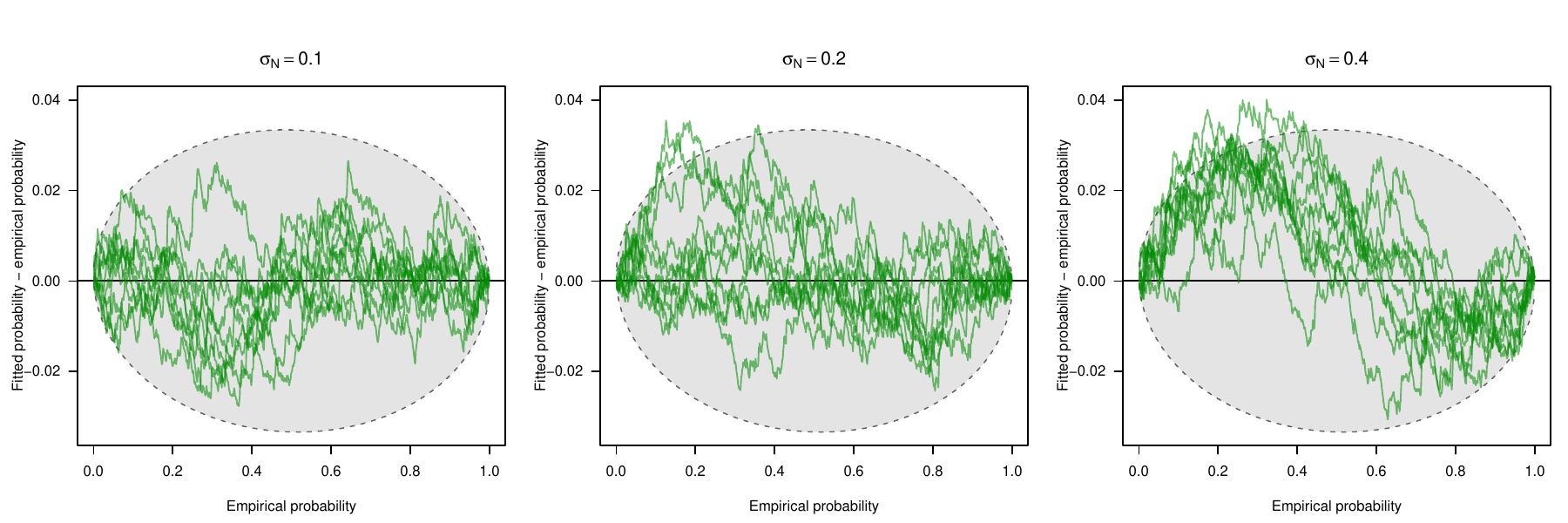}
\caption{
PP difference plots, i.e., $(p_i, \widehat{F}_{Y_{\operatorname{GPD}}}(Y_{(i)}) - p_i,)$ for $i=1,\ldots, m_{u+\delta}$, for na\"ive GPD fits to the noisy observations
$Y_i$ with $\xi=-0.15$. Panels correspond to
$\sigma_N=0.1,0.2,0.4$. Green curves show ten simulation replicates,
and the grey region gives the pointwise 95\% tolerance intervals.}
\label{fig:sim_case1_naive_qq}
\end{figure}

Figure~\ref{fig:sim_case1_tail_params} has shown that using the GPD-ME can lead to substantial improvements over the na\"ive method. But, can we easily assess whether the GPD-ME may improve the fit without performing the inference? To help assess this, in Figure~\ref{fig:sim_case1_naive_qq}, we show a measure of the goodness-of-fit from using the na\"ive method. Explicitly, it shows PP difference plots for na\"ive GPD fits to the noisy exceedances, i.e., the $Y_i>u+\delta$, using ten randomly selected simulation replicates for each measurement error scale. Let $\widehat F_{Y_{\operatorname{GPD}}}$ denote the fitted GPD distribution
function for the noisy exceedances. For each
replicate, let $Y_{(i)}$, $i=1,\ldots,m_{u+\delta}$, denote the ordered (increasing with $i$) noisy
exceedances, where $m_{u+\delta}$ is the number of noisy exceedances above the
threshold. We compare the fitted probabilities
$\widehat{F}_{Y_{\operatorname{GPD}}}(Y_{(i)})$ with the empirical probabilities
$p_i=i/(m_{u+\delta}+1)$. The
plotted difference is
$
\widehat{F}_{Y_{\operatorname{GPD}}}(Y_{(i)}) - p_i,$ for $i=1,\ldots,m_{u+\delta}.
$ Values close to zero indicate good agreement between the empirical probabilities and the fitted tail model. When $\sigma_N=0.1$, the PP difference curves remain close to zero and
mostly lie within the pointwise 95\% tolerance intervals, so the standard PP diagnostic does
not provide much evidence against the na\"ive GPD model. A similar
conclusion holds for $\sigma_N=0.2$, although the departures from zero
are slightly larger, with a few of the replications now giving values outside these intervals. The departure is clearer still when $\sigma_N=0.4$,
where the curves tend to be positive for lower empirical probabilities
and negative for larger empirical probabilities. The corresponding QQ plots can be found in supplementary material. To summarise this diagnostic across all 200 simulation replicates, we
record the proportion of replicates in which the PP difference curve
falls outside the pointwise 95\% tolerance intervals at least once. These
proportions are $0.46$, $0.65$ and $0.84$ for
$\sigma_N=0.1$, $0.2$ and $0.4$, respectively. This increase shows
that data departures from the na\"ive GPD fit become more common as the
measurement error scale increases, and hence that the evidence to use GPD-ME over the na\"ive method increases with $\sigma_N$.

\begin{figure}[H]
\centering
\includegraphics[width=0.65\textwidth]{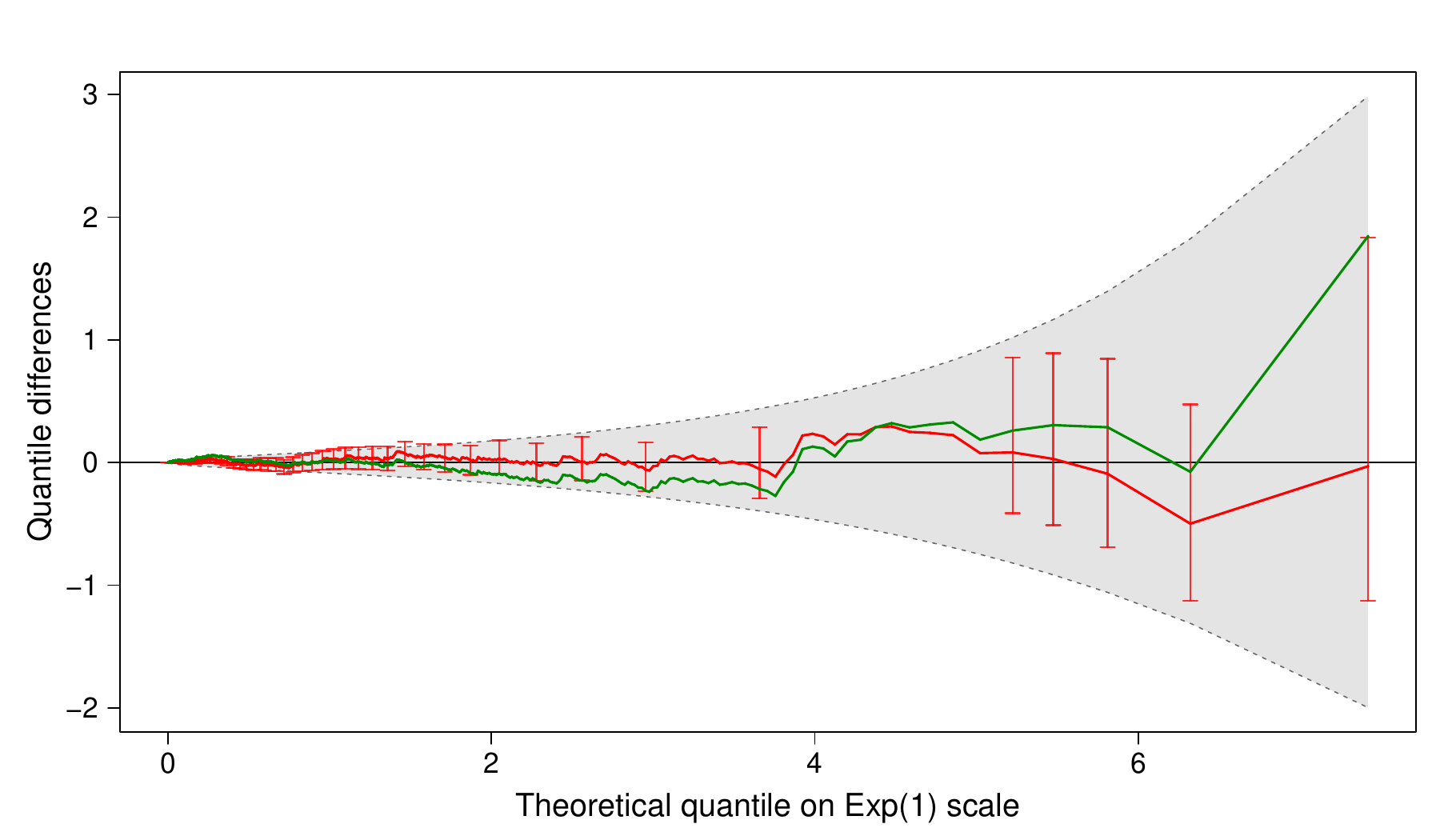}
\caption{
QQ-difference plot on standard exponential margins for scenario 1 with
$\sigma_N=0.4$. The GPD-ME curve (red) and the na\"ive GPD curve (green)
show the pointwise median QQ differences across simulation replicates.
The grey region gives pointwise 95\% tolerance intervals, and the red vertical bars give pointwise 95\% (over replicates fits)
intervals for the GPD-ME curve.
}
\label{fig:sim_case1_sigma04_qqdiff}
\end{figure}

To examine the high-noise case more directly,
Figure~\ref{fig:sim_case1_sigma04_qqdiff} compares the GPD-ME and
na\"ive GPD fits for the simulation setting with $\sigma_N=0.4$. For each simulation replicate, both models were fitted to the same noisy
dataset. Following the strategy of \citet{heffernan2001extreme}, the observed exceedances
$Y_i>u+\delta$ were transformed to standard exponential margins using the probability integral transform $
U_i
=
-\log\{1-\widehat{F}(Y_i)\},
$
where $\widehat F$ denotes the fitted conditional distribution function for exceedances above $u+\delta$ under the
corresponding model. The
ordered transformed values $U_{(i)}$ are then compared with the empirical probabilities through the theoretical
exponential quantiles $Q_{\mathrm{Exp}(1)}(p_i)$. The vertical axis in
Figure~\ref{fig:sim_case1_sigma04_qqdiff} shows the QQ difference, i.e., 
$
U_{(i)} - Q_{\mathrm{Exp}(1)}(p_i),
$
so values close to zero indicate good agreement with the fitted tail
model. 

The QQ-difference plot shows that the  pointwise median curve (over replicate fits) for the na\"ive GPD fit has a more consistent deviation from the exponential reference than the GPD-ME curve. The median na\"ive curve is slightly below zero over much of the lower and middle
quantile range, but then becomes positive in the upper
tail. In contrast, the median GPD-ME curve remains closer to zero over most of the range, and the pointwise 95\% confidence intervals include zero at all shown quantiles.

We now investigate more fully the relative performance of the GPD-ME and
na\"ive methods observed in Figure~\ref{fig:sim_case1_tail_params}. Since the oracle method is the GPD fitted to the latent variables $X_i$, it is useful as the benchmark. If RMSE provides a useful summary for comparing the three methods, we would expect the ordering $
\operatorname{RMSE}_{\mathrm{Oracle}}
\leq
\operatorname{RMSE}_{\mathrm{GPD\mbox{-}ME}}
\leq
\operatorname{RMSE}_{\text{Na\"ive}}$.
Therefore, we summarise the comparison between the GPD-ME and na\"ive methods using the relative RMSE efficiency, defined as
$$E_{\mathrm{RMSE}} =
(\operatorname{RMSE}_{\mathrm{GPD\mbox{-}ME}} -
\operatorname{RMSE}_{\mathrm{Oracle}})
/
(\operatorname{RMSE}_{\mathrm{\text{Na\"ive}}}
-
\operatorname{RMSE}_{\mathrm{Oracle}}),$$ with values given in Table~\ref{tab:rmse_efficiency}. Under the expected ordering, $E_{\mathrm{RMSE}}=0$ corresponds to the GPD-ME having the oracle-level of RMSE, while
$E_{\mathrm{RMSE}}=1$ corresponds to it having the na\"ive RMSE. Values of $E_{\mathrm{RMSE}}$ between
zero and one indicate the relative level that the GPD-ME estimator improves on the na\"ive
method. Values of $E_{\mathrm{RMSE}}$ greater than one
indicate that the GPD-ME estimator does not improve on the na\"ive method in terms of RMSE. Negative values of $E_{\mathrm{RMSE}}$ occur when either the na\"ive or GPD-ME estimator has a smaller empirical RMSE than the oracle estimator.
Such cases should be interpreted cautiously, since they may reflect the limitation of $\mathrm{RMSE}$ as a reliable metric of inference performance rather than showing a meaningful improvement over the oracle benchmark.

For moderate and large noise levels, $E_{\mathrm{RMSE}}$ values for both of $(\xi, \sigma_{u+\delta})$ parameters are between zero and one and decrease as $\sigma_N$ increases. This
indicates that the GPD-ME estimator becomes closer to the oracle RMSE,
relative to the na\"ive RMSE, when measurement error has a stronger
effect. In the $\sigma_N=0.1$ case, the $E_{\mathrm{RMSE}}$ measure is less informative at least for this choice of the sample size $n$, because the measurement error is weak and the
empirical RMSEs of the three methods are very similar, with surprisingly $\mathrm{RMSE}_{\text{Na\"ive}}< \mathrm{RMSE}_{\text{Oracle}}$. In this setting,
small differences in RMSE can make the relative efficiency measure appear
unstable, producing values outside the interval $[0,1]$. The similar issue also occurs in the $\xi=0.15$ sensitivity results reported in the
supplementary material. Thus, when the signal-to-noise ratio is large, explicitly modelling the noise does not necessarily improve RMSE, since the additional uncertainty from estimating the measurement error component can outweigh the small reduction in bias. 
This interpretation is supported by the PP diagnostics in
Figure~\ref{fig:sim_case1_naive_qq}. For $\sigma_N=0.1$, the PP difference curves provide little evidence
against the na\"ive GPD fit, which agrees with the RMSE comparison in this high signal-to-noise
setting. The bias and variance components underlying these RMSE values are reported
in the supplementary material.

\begin{table}[t]
\centering
\caption{
Relative RMSE efficiency $E_{\mathrm{RMSE}}$ of the GPD-ME estimator for the latent variable of GPD parameters over a range of signal-to-noise ratios in scenario 1 with $\xi=-0.15$. Efficiency evaluated using two different KDEs for the body in the GPD-ME method: (i) a fixed KDE based on the latent variables $\boldsymbol{X}$ and (ii) a KDE estimated from the noisy observations $\boldsymbol{Y}$.
}
\label{tab:rmse_efficiency}

\setlength{\tabcolsep}{9pt}
\renewcommand{\arraystretch}{1.25}

\begin{tabular}{@{}llccc@{}}
\hline
Quantity &
KDE used in GPD-ME &
$\sigma_N=0.1$ &
$\sigma_N=0.2$ &
$\sigma_N=0.4$ \\
\hline

\multirow{2}{*}{$E_{\mathrm{RMSE}}(\hat{\xi})$}
& $X$-based KDE & $-2.67$ & 0.89 & 0.74 \\
& $Y$-based KDE & $-3.06$ & 0.94 & 0.77 \\

\hline

\multirow{2}{*}{$E_{\mathrm{RMSE}}(\widehat{\sigma}_{u+\delta})$}
& $X$-based KDE & 1.38 & 0.70 & 0.49 \\
& $Y$-based KDE & 0.92 & 0.66 & 0.53 \\

\hline
\end{tabular}

\end{table}
As a sensitivity analysis of the results in Table~\ref{tab:rmse_efficiency}, we replace the fixed body KDE based on the latent variables $\boldsymbol{X}$ by a KDE estimated from the noisy observations $\boldsymbol{Y}$. This is closer to the practical setting, where the latent body distribution is not observed
directly. The corresponding RMSE efficiency values are also shown in Table~\ref{tab:rmse_efficiency}. The results are generally similar (approximately equal numbers of improved and weaker performances) to those obtained using the $X$-based KDE, indicating that the main conclusions are not driven by the idealised body estimate. This supports the view that the main benefit of
the proposed approach comes from modelling measurement error in the tail
inference, while the partially censored likelihood reduces sensitivity to
the precise form of the body estimate below the threshold. Further parameter estimate boxplots for this
sensitivity analysis are reported in the supplementary material. For brevity, this sensitivity analysis is not repeated in the later
covariate-dependent scenarios, where similar qualitative patterns were
observed.

\begin{figure}[H] 
\centering \includegraphics[width=0.52\textwidth]{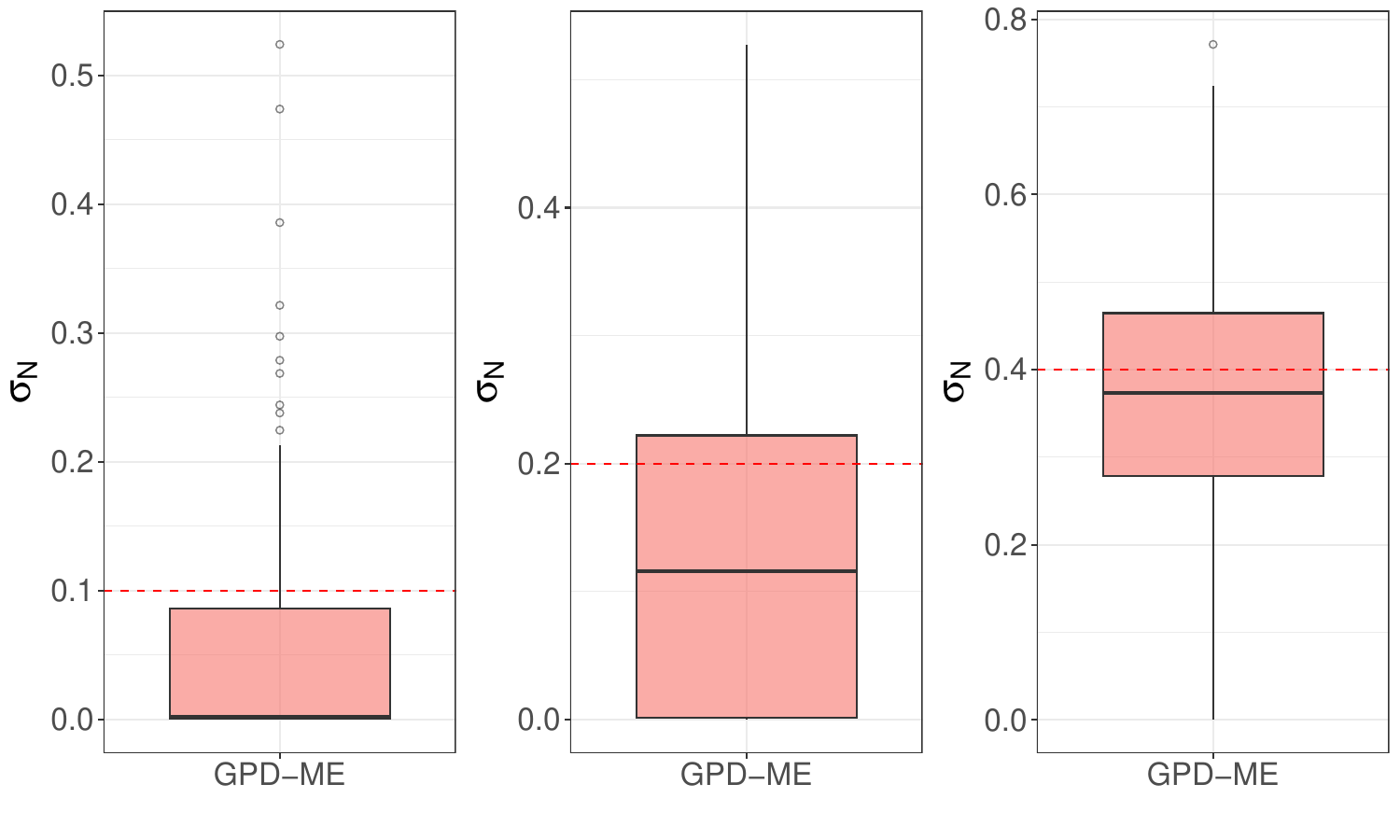} 
\caption{ Sampling distributions of the GPD-ME estimates of the measurement error scale $\sigma_N$ in scenario 1 for $\sigma_N=0.1, 0.2, 0.4$ (left to right). Dashed horizontal lines indicate the true noise standard deviation levels. } 
\label{fig:sim_case1_sigmaN} 
\end{figure}

Figure~\ref{fig:sim_case1_sigmaN} shows the corresponding sampling distribution of the measurement error scale estimates for the same three levels of $\sigma_N$ using the GPD-ME method. When the signal-to-noise ratio is large,
corresponding to $\sigma_N=0.1$, the majority of estimates of $\sigma_N$ are close to zero with a negative bias, indicating that very small measurement error is difficult to distinguish from variability in the latent variable distribution. As the true noise level increases, the measurement error scale becomes more identifiable. For $\sigma_N=0.2$, the estimates remain variable and negatively biased but they are shifted away from zero, while for $\sigma_N=0.4$ they are more clearly centred around the true value. This reflects the increasing information about the measurement error when the noise contribution is larger. These results in Figure~\ref{fig:sim_case1_sigmaN} potentially could be improved by incorporating the penalty $P_{N}(\theta_{N})$ of expression~\eqref{eq:penalised_likelihood} if this has the property that $P_{N}(\theta_{N})\rightarrow0$ as $\sigma_{N}\rightarrow0$. This point is particularly relevant as a user is not likely to use the GPD-ME method unless measurement error is anticipated or that diagnostics such as in Figure~\ref{fig:sim_case1_naive_qq}, suggested that $\sigma_{N}\neq0$.

We also examine the effect of measurement error on high-quantile
inference by calculating $100$-year and $475$-year return levels from each
fitted tail model. The $475$-year return level is included because it is a
standard return period in seismic hazard analysis and was also used for
the Groningen application in \citet{yue2025integrating}. Let $m_{u+\delta}(y)$ denote the annual rate of exceedances
above $u+\delta$. For a return period $T$, the return level $a_T$
is defined as the level expected to be exceeded once every
$T$ years. Under the fitted GPD tail model with scale
$\sigma_{u+\delta}$ and shape $\xi$, $a_T$ is given by
\[
a_T
=
\begin{cases}
u+\delta
+
\dfrac{\sigma_{u+\delta}}{\xi}
\left\{(T m_{u+\delta}(y))^{\xi}-1\right\},
& \xi\ne 0, \\[12pt]
u+\delta
+
\sigma_{u+\delta}\log(Tm_{u+\delta}(y)),
& \xi=0 ,
\end{cases}
\]
where we take $m_{u+\delta}(y)=54$, chosen to match the largest annual exceedance count used in the Groningen application. The return level results are reported in the supplementary material (Table~S2). They show a different pattern from the individual tail parameter results. Although the na\"ive GPD fit gives biased estimates of both $\xi$ and $\sigma_{u+\delta}$, these biases can partially offset when converted into return levels. As a result, the na\"ive method can have smaller RMSE for some return levels, despite giving less accurate tail-parameter estimates. For the GPD-ME method, the return level bias remains small across the three noise levels, but the variance increases when $\sigma_N=0.4$, leading to larger RMSE for both the $100$-year and $475$-year return levels.

These slightly counter-intuitive findings using RMSE reflect some limitation of RMSE as a performance metric for inference. Specifically,
RMSE measures point-estimation accuracy, but it does not assess whether
the uncertainty intervals are reliable. For extreme value inference, this
is particularly important because conclusions depend not only on point
estimates of the shape parameter and high quantiles, but also on the
confidence intervals attached to these estimates. A method can have a small RMSE if
its point estimates are close to the truth on average, while still giving
confidence intervals that are too narrow or poorly centred. Therefore, we
also examine the coverage of the GPD-ME and na\"ive methods, which records the proportion of simulated datasets
for which the estimated bootstrap-based 95\% confidence interval contains the true value.
Table~\ref{tab:case1_coverage_all} reports the empirical coverage for
$\xi$ and the conditional GPD 90\% and 99.5\% quantiles, i.e., quantiles of
the fitted exceedance distribution above $u+\delta$, conditional on the
threshold being exceeded. The 95\% confidence intervals for the empirical coverage probabilities
are also reported, with these being obtained by treating the number of covered simulation replicates as a
binomial count.

\begin{table}[t]
\centering
\caption{
Empirical coverage of 95\% bootstrap-based confidence intervals in scenario 1, based on 1000 simulated datasets. The reported intervals are
95\% binomial confidence intervals based on the corresponding
coverage counts.
}
\label{tab:case1_coverage_all}

\setlength{\tabcolsep}{6pt}
\renewcommand{\arraystretch}{1.15}
\resizebox{0.85\textwidth}{!}{%
\begin{tabular}{@{}llcccccc@{}}
\hline
& &
\multicolumn{2}{c}{$\sigma_N=0.1$} &
\multicolumn{2}{c}{$\sigma_N=0.2$} &
\multicolumn{2}{c@{}}{$\sigma_N=0.4$} \\
\cline{3-4}
\cline{5-6}
\cline{7-8}
Quantity &
Method &
Coverage &
Interval &
Coverage &
Interval &
Coverage &
Interval \\
\hline

\multicolumn{8}{@{}l}{\textit{Shape parameter $\xi$}} \\
&
GPD-ME &
0.926 &
$[0.908,\,0.941]$ &
0.943 &
$[0.927,\,0.957]$ &
0.894 &
$[0.873,\,0.912]$ \\
&
Na\"ive &
0.892 &
$[0.871,\,0.911]$ &
0.691 &
$[0.661,\,0.720]$ &
0.270 &
$[0.243,\,0.299]$ \\
&
Oracle &
0.947 &
$[0.931,\,0.960]$ &
0.947 &
$[0.931,\,0.960]$ &
0.947 &
$[0.931,\,0.960]$ \\[4pt]

\multicolumn{8}{@{}l}{\textit{Conditional GPD 90\% quantile}} \\
&
GPD-ME &
0.933 &
$[0.916,\,0.948]$ &
0.972 &
$[0.960,\,0.981]$ &
0.925 &
$[0.907,\,0.941]$ \\
&
Na\"ive &
0.931 &
$[0.913,\,0.946]$ &
0.674 &
$[0.644,\,0.703]$ &
0.010 &
$[0.005,\,0.018]$ \\
&
Oracle &
0.941 &
$[0.925,\,0.955]$ &
0.941 &
$[0.925,\,0.955]$ &
0.941 &
$[0.925,\,0.955]$ \\[4pt]

\multicolumn{8}{@{}l}{\textit{Conditional GPD 99.5\% quantile}} \\
&
GPD-ME &
0.933 &
$[0.916,\,0.948]$ &
0.926 &
$[0.908,\,0.941]$ &
0.900 &
$[0.880,\,0.918]$ \\
&
Na\"ive &
0.862 &
$[0.839,\,0.883]$ &
0.893 &
$[0.872,\,0.911]$ &
0.471 &
$[0.440,\,0.502]$ \\
&
Oracle &
0.946 &
$[0.930,\,0.959]$ &
0.946 &
$[0.930,\,0.959]$ &
0.946 &
$[0.930,\,0.959]$ \\
\hline
\end{tabular}}

\end{table}

Table~\ref{tab:case1_coverage_all} shows that for $\sigma_N=0.1$, all three inference methods give coverage close to the nominal 95\% level for all three quantities. As the signal-to-noise decreases, the coverage of the na\"ive method rapidly deteriorates, particularly for the shape parameter and the 90\% quantile. For example, when $\sigma_N=0.4$, the na\"ive coverage falls to 0.27 for $\xi$ and 0.01 for the conditional GPD 90\% quantile. The deterioration is less severe for the
GPD 99.5\% quantile, which may reflect partial compensation between the
na\"ive method's overestimation of $\sigma_{u+\delta}$ and underestimation
of $\xi$ for into very high quantiles. However, this
compensation breaks down as the measurement error becomes large, with the
na\"ive method's coverage falling to 0.470 when $\sigma_N=0.4$. In contrast, the GPD-ME method maintains a coverage, much closer to the
nominal level, across all three noise levels and with coverage very close to the oracle method. For $\xi$, the GPD-ME
coverage ranges from 0.89 to 0.94, while for the conditional GPD 90\% and 99.5\%
quantiles it remains between 0.90 and 0.97. Overall, these results show that ignoring
measurement error can lead to substantial undercoverage for both tail
parameters and conditional tail quantiles, especially when the
measurement error scale is large. Accounting explicitly for observation error gives more reliable
uncertainty quantification, and that the GPD-ME method achieves
excellent coverage despite the presence of measurement noise.

\subsection{Scenario 2: Covariate-dependent scales}
Scenario 2 extends the constant-scale setting by allowing both the latent GPD scale and the measurement error scale to depend on covariates. The scale functions $g_L$ and $g_N$ for the latent GPD distribution and the measurement error distribution follow the specifications used in the Groningen application, see Section~\ref{subsec:covariates_app} for details of the covariates and their construction. For each simulation replicate, new values of the latent scale covariate $Z_L$ and measurement error covariate $Z_N$ are generated by sampling independently from their empirical marginal distributions, hence $(Z_L, Z_N)$ are independent here, unlike scenario 3 or the Groningen application in Section~\ref{Sec: Application}. The two covariates are sampled independently in each replicate.

\begin{figure}[H]
    \centering
    \includegraphics[width=0.82\linewidth]{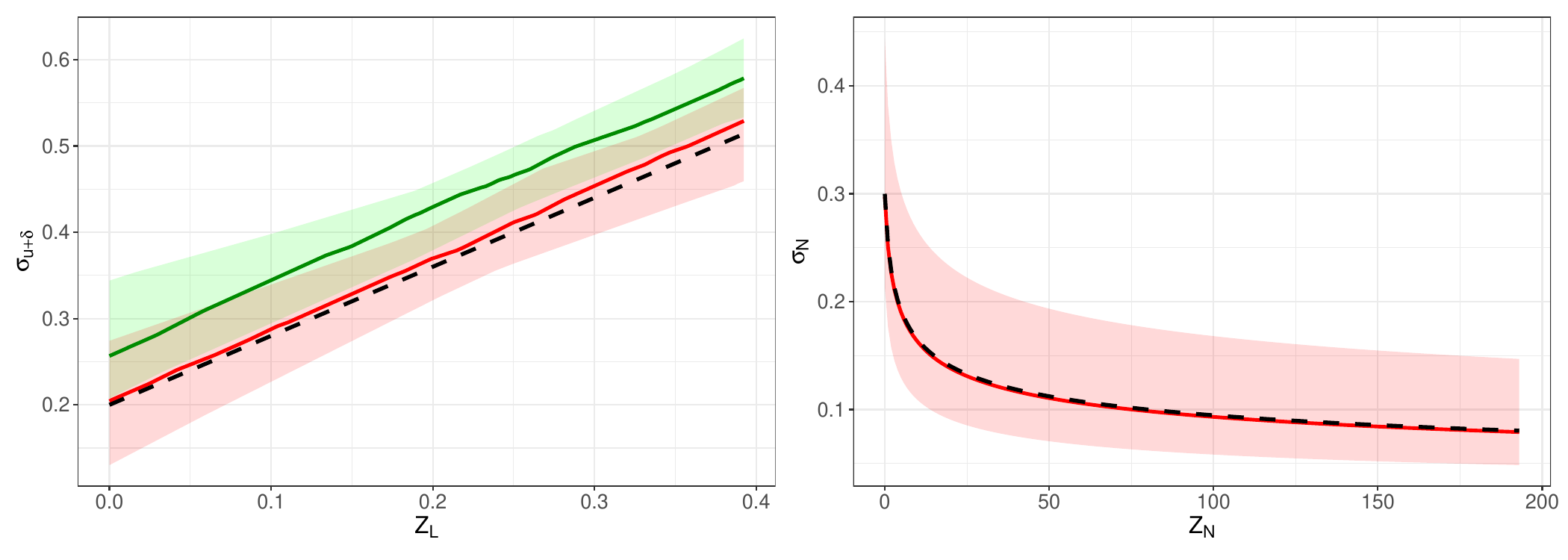}
\caption{Estimated covariate effects in simulation scenario 2 for GPD-ME (in red) and na\"ive (in green) methods. Left: fitted GPD scale as a function of $Z_L$. Right: fitted measurement error scale as a function of $Z_N$ under the GPD-ME model. Dashed black lines show the true curves. Solid line and shaded bands show the sampling medians and 95\% confidence intervals respectively for the
corresponding fitted curves.}
\label{fig:sim_case2_covariates}
\end{figure}

For each simulated sample, the true parameters are $\xi=-0.15,\,
\boldsymbol{\theta}_L=(\beta_0,\beta_1)=(0.2,0.8),\\\,
\boldsymbol{\theta}_N=(\sigma_{N_0},\gamma)=(0.3,0.25),
$
thus, we have $\sigma_{u+\delta} \in (0.2,0.51)$ and $\sigma_N \in (0.08,0.3)$ over the simulated covariate ranges and with the functions for $\sigma_{u+\delta}$ and $\sigma_{N}$ shown in Figure~\ref{fig:sim_case2_covariates} and given mathematically in expressions~\eqref{sigmaGPD} and~\eqref{eq:noisescale} respectively. We compare the proposed GPD-ME method with a na\"ive GPD fit to the noisy observations, which uses the same covariate structure formulation for the GPD scale parameter but it ignores measurement error. For this covariate-based inference, we use the discretised likelihood approximation of Section~\ref{subsec:covariate_discretisation}.
Both the latent scale covariate $Z_L$ and the measurement error covariate $Z_N$ are divided into $G_L=G_N=6$ bins, with the median value in each bin used as the representative covariate value, $Z_{L,r}^{\mathrm{rep}}$ and $Z_{N,s}^{\mathrm{rep}}$, for $r,s=1,2,\ldots,6$, respectively. The choice of six bins provides a balance between numerical resolution
and computational efficiency: it allows the fitted scales to vary across
the covariate ranges while limiting the number of numerical convolution evaluations required by the
FFT implementation.

Figure~\ref{fig:sim_case2_covariates} shows the fitted covariate effects
across the simulation replicates. The left panel compares the fitted
latent GPD scale function ($\sigma_{u+\delta}$) as a function of the latent scale covariate $Z_L$ for GPD-ME and na\"ive method. The GPD-ME
fit follows the true curve closely, whereas the na\"ive fit overestimates over much of the covariate
range. As in scenario~1, this illustrates that when measurement error is ignored, part of
the variability induced by observation error is absorbed into the
estimated GPD scale. Figure~\ref{fig:sim_case2_covariates} (right) shows the fitted measurement error
scale from the GPD-ME model performs well with the bootstrap median closely following the
true decreasing relationship between the $\sigma_N$ and $Z_N$.

Table~\ref{tab:sim_case2_xi} summarises estimation of the GPD shape
parameter $\xi$. The GPD-ME method gives a smaller bias and RMSE than the na\"ive method. As in the iid case,
ignoring measurement error leads to a larger bias in the na\"ive method estimated shape
parameter than the GPD-ME method. Both empirical 95\% sampling intervals contain the true value, but the
GPD-ME interval is slightly wider, reflecting the additional uncertainty
from estimating the measurement error component. Compared with the small-noise case in scenario~1, the RMSE results in scenario~2 show a clearer benefit from modelling measurement error. Overall, scenario~2 shows that the proposed method can recover the main
covariate effects when the latent scale and measurement error covariates
vary independently. By sampling $Z_L$ and $Z_N$ independently, this scenario avoids issues of confounding that can
arise when the two scale components are driven by correlated covariates. Scenario~3, in the
supplementary material, considers the more difficult case where the two
covariates are sampled jointly from the Groningen catalogue. In that setting, the performance of the standard GPD-ME fit
deteriorates relative to the independent covariate case, particularly in
the recovery of the covariate effects. This
reflects the additional difficulty of separating variation in the latent
GPD scale from variation in the measurement error scale when the two
covariates are correlated. The regularised GPD-ME
fit, introduced in Section~\ref{subsec:penalisation}, gives more stable recovery of the covariate effects in this setting.

\begin{table}[t]
\centering
\caption{
Simulation summary for the GPD shape parameter $\xi$ inference in scenario 2,
based on 200 simulated datasets. The interval column presents the 95\% sampling
distribution interval for estimates of $\xi$.
}
\label{tab:sim_case2_xi}

\setlength{\tabcolsep}{10pt}
\renewcommand{\arraystretch}{1.15}

\begin{tabular}{@{}lccc@{}}
\hline
Method & Bias & Interval & RMSE \\
\hline
GPD-ME & $-0.010$ & $[-0.233,\,-0.097]$ & 0.039 \\
Na\"ive & $-0.042$ & $[-0.241,\,-0.131]$ & 0.052 \\
\hline
\end{tabular}

\end{table}

\section{Application: Groningen Induced Earthquake Magnitudes}
\label{Sec: Application}
\subsection{Earthquake catalogue}

\begin{figure}[H]
    \centering
    \includegraphics[width=0.85\linewidth]{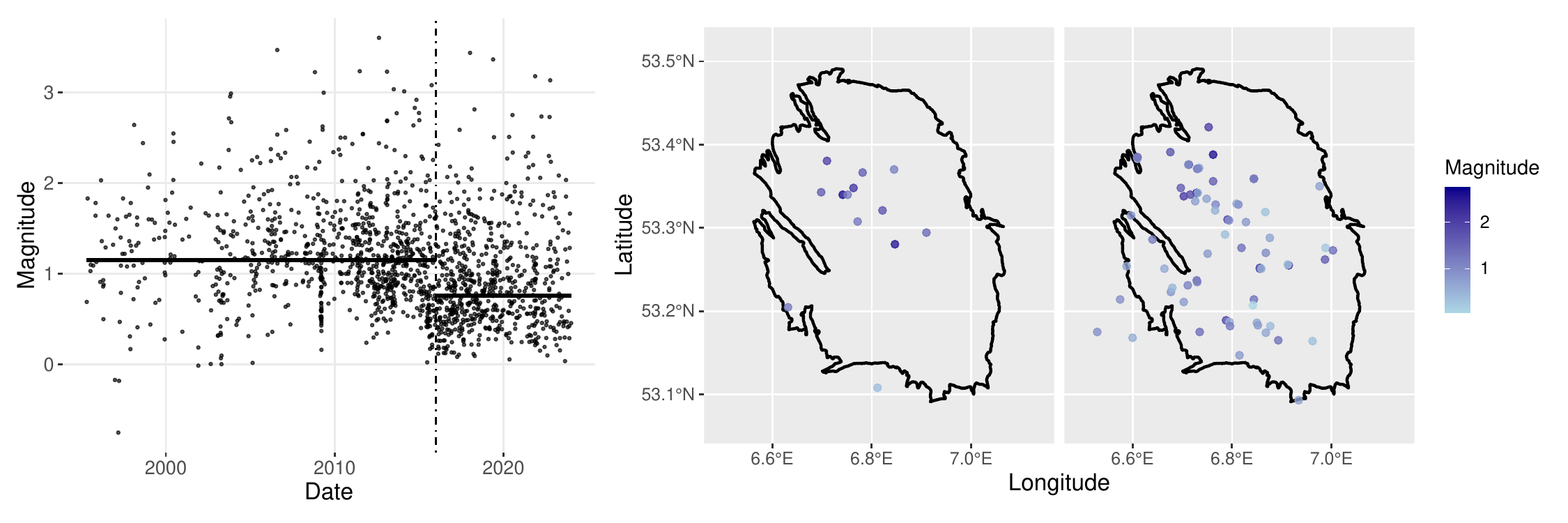}
    \caption{Groningen earthquake catalogue. Left: recorded local magnitudes,
$M_L$, over time, with horizontal lines showing the thresholds used by
\citet{varty2021inference}, and a vertical line marking the 2016 change
in GPD threshold. Right: spatial event locations in 2010 and
2020 respectively, coloured by magnitude in $M_L$, with the gas field boundary presented with black line.}
\label{fig:groningen_summary}
\end{figure}

We apply the proposed GPD-ME methodology to the Groningen earthquake catalogue.
The $900$km$^2$ Groningen gas field, shown in Figure~\ref{fig:groningen_summary} (right) is located in the north of the Netherlands. It is Europe's largest gas field and has experienced induced seismicity
associated with gas extraction since the early 1990s. Although the individual events are moderate in magnitude, their
occurrence at shallow depths, typically around $3$--$4$ km, can
result in substantial localised damage. Furthermore, the proximity of
the Groningen gas field to populated areas makes the catalogue an
important case study for seismic risk assessment. The Groningen earthquake catalogue is also of relatively high quality, due to 
major investment in
the monitoring and measurement network, making it particularly suitable
for statistical modelling of induced seismicity. Nevertheless, careful treatment of measurement uncertainty remains necessary. Understanding the
magnitude distribution, particularly its upper tail, is important for
estimating the probability of larger events and can inform risk
assessment and mitigation planning. For background detail of this gas field see \cite{beirlant2019estimating}.

To remain consistent with previous analyses of the Groningen catalogue,
we use the same study period as \citet{yue2025integrating} and \citet{Murphy2026spatio}. The catalogue used in this analysis covers
the period from April 1995 to January 2024 and contains $n=1565$
seismic events within the Groningen region of interest. For each event $i=1,\ldots,n$, the catalogue
contains the event time $t_i$, hypocentral location $\boldsymbol{s}_i$,
and recorded local magnitude $y_i$. The hypocentre
$\boldsymbol{s}_i$ consists of a two-dimensional surface location,
reported in the Dutch RD coordinate system. The recorded magnitudes are given on the local
magnitude scale $M_L$, a logarithmic scale for earthquake size. Earthquake magnitudes are not directly observed quantities. Instead,
they are inferred from seismic recordings made by sensors, referred to
here as geophones, and are therefore subject to
measurement uncertainty. In our notation when an earthquake is detected, the recorded magnitude $y_i$
is treated as a noisy measurement of a latent magnitude $x_i$. This is
the central motivation for applying the measurement error framework
developed in Sections~\ref{Sec:Modelsetup}--\ref{Sec:Inference} to
the Groningen catalogue.

The catalogue contains an increasing number of small-magnitude events
over time, reflecting changes in the detection capability of the geophone
monitoring network. This is usually described through the magnitude of
completeness, $m_c(t)$, defined as the magnitude above which events are
assumed to be certain to be recorded at time $t$. Earlier work used a
conservative constant threshold $m_c(t)=1.45\,M_L$ for Groningen \citep{dost2013august, beirlant2019estimating}, while later analyses
allowed $m_c(t)$ to vary over time to reflect improvements in the
monitoring network and the increased ability to detect smaller events
\citep{varty2021inference,yue2025integrating}. More recently, \citet{Murphy2026spatio} allowed the magnitude of
completeness to vary spatio-temporally, using the evolving spread of the
geophone network to capture variation in catalogue detectability. Although the magnitude of completeness concerns the detection of events rather than the
accuracy of their recorded magnitudes, both are affected by the
configuration of the monitoring network. The variation in
$m_c(t)$ therefore supports the broader view that detection capability changes over the catalogue period, motivating
event-specific modelling of measurement uncertainty.

In the present analysis, we set the lower threshold in the latent
body--tail model to $u=0.76\,M_L$. This corresponds to the lower
threshold estimated by \citet{varty2021inference}, whose
threshold-detection procedure found the change in catalogue completeness
around 2016. This timing coincides with the post-network-expansion
period, when the improved monitoring network allowed smaller events to
be recorded more reliably, see Figure~\ref{fig:groningen_summary_geo} (left). We take
$\delta=0.4$, so that the GPD tail component is used fully above
$u+\delta=1.16\,M_L$. This upper value is close to the threshold, before the geophone network expansion, as estimated by \citet{varty2021inference}. The interval
$(0.76,1.16)$ provides a transition region reflecting uncertainty
about where the latent GPD tail begins to dominate. For the na\"ive GPD model, we use the threshold $u+\delta$. Figure~\ref{fig:groningen_summary} summarises the catalogue. The left panel shows the recorded magnitudes over time with the
horizontal lines indicating the thresholds used by \cite{varty2021inference} for modelling the upper tail of the magnitude
distribution. The right panel shows the spatial distribution of catalogue events recorded
in 2010 and 2020, with points coloured by recorded local magnitude
$M_L$. The events are concentrated within the gas field region, with a notable difference in the number and spatial variation in event locations and magnitudes between the
two years.

\subsection{Covariates}
\label{subsec:covariates_app}
\subsubsection{Latent covariate: Kaiser stress}
\label{subsubsec:groningen_stress}

The latent earthquake magnitude marginal distribution is known to vary with the physical
state of the reservoir. Gas extraction changes the subsurface stress
field, and stress-related quantities have previously been used to explain
spatial and temporal variation in induced seismicity in Groningen
\citep{bourne2020stress,Murphy2026spatio}. Therefore, we use the same stress-related covariate, Kaiser stress, in the latent hybrid density model~\eqref{eq:latentmodel}, as it provides a physically motivated covariate that can be used as a proxy for the evolving stress state in the reservoir.

\begin{figure}[H]
    \centering
    \includegraphics[width=0.89\linewidth]{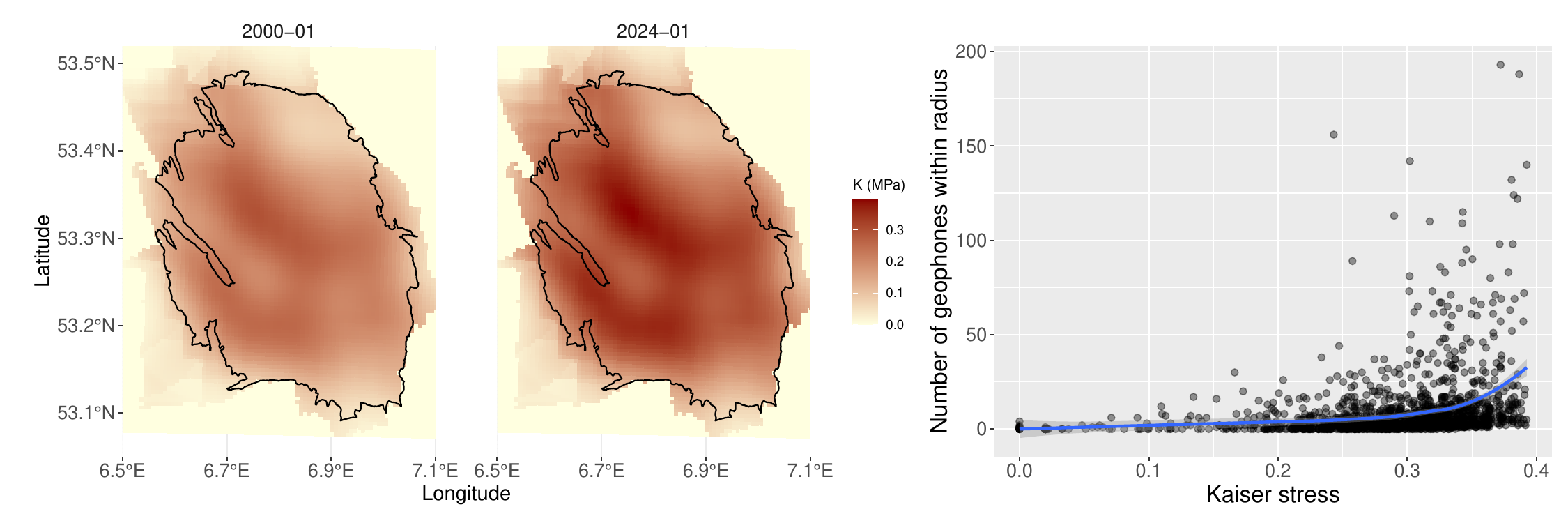}
    \caption{
    Covariates used in the Groningen application. Left: Kaiser stress
fields ($\mathrm{MPa}$) for January 2000 and January 2024. Right: scatter plot of the number
of detecting geophones, $N_{g,i}$, within the detection radius of the $i$-th earthquake, against the corresponding Kaiser
stress value, $K_i$, for each recorded event. The blue curve shows a locally smoothed trend fitted to the scatter
points.}
    \label{fig:groningen_covariates}
\end{figure}

We denote the Kaiser stress field by $K(\boldsymbol{s},t)$, where
$K(\boldsymbol{s},t)$ represents the maximum change in stress
experienced at location $\boldsymbol{s}$ up to time $t$, relative to
the initial stress state at the beginning of gas extraction. Hence for each $\boldsymbol{s}$, $K(\boldsymbol{s},t)$ is a non-decreasing function of $t$. Kaiser stress is derived from computationally intensive numerical geophysical reservoir models driven by the history of gas
extraction and the associated changes in the subsurface stress field \citep{bourne2020stress}. The Kaiser stress field is available on a spatial grid ($0.5\,\mathrm{km}\times
0.5\,\mathrm{km}$) over the
Groningen region and on a monthly time scale across the temporal scale of the data catalogue. For each event $i$, we assign an
event-specific stress value
\[
Z_{L,i} := K_i = K(\boldsymbol{s}_i,t_i),
\]
using the value of the Kaiser stress field at the grid cell in the numerical model grid which contains the event hypocentre $\boldsymbol{s}_i$ and at the corresponding event
time $t_i$ and we treat it as
constant within each month. Figure~\ref{fig:groningen_covariates} (left) illustrates
the spatio-temporal variation in the Kaiser stress field for January
in 2000 and in 2024. The spatial pattern is non-uniform, and the
overall stress level increases over time, reflecting the evolving
physical state of the reservoir under continued gas extraction over this period and with a spatial pattern that is consistent with earthquake locations shown in Figure~\ref{fig:groningen_summary}.

\begin{figure}[H]
    \centering
    \includegraphics[width=0.85\linewidth]{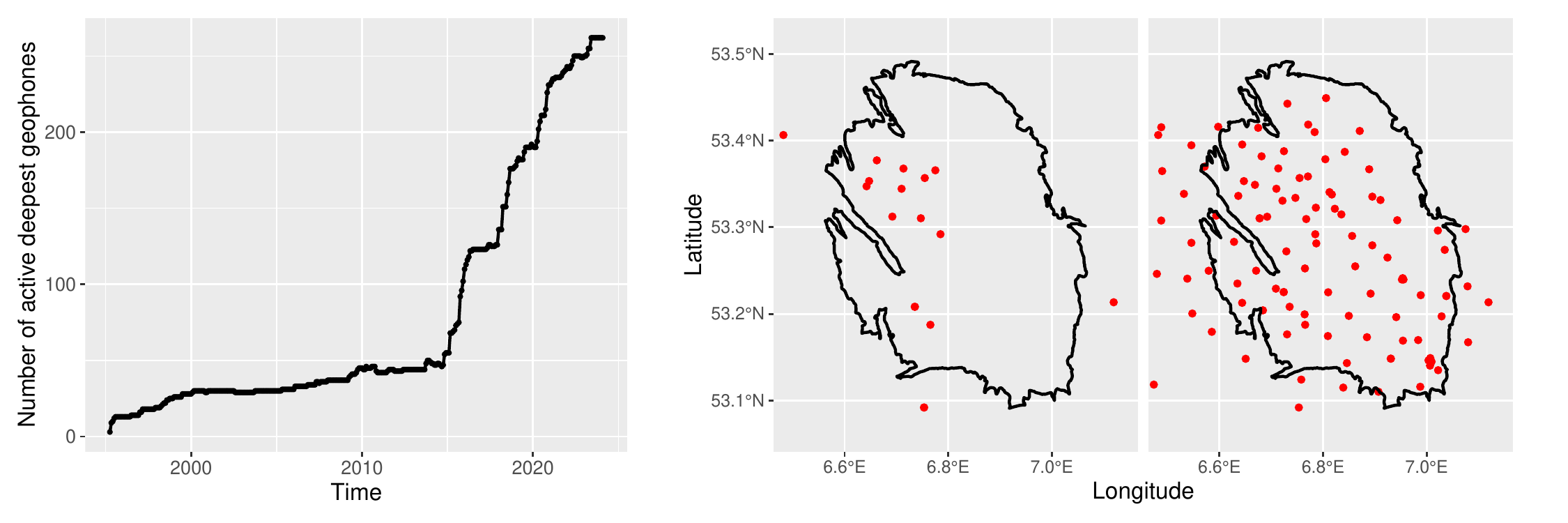}
\caption{
Geophone network catalogue: Number of active geophones over time (left) and spatial distribution of geophones in 2010 (middle) and 2020 (right).}
\label{fig:groningen_summary_geo}
\end{figure}

In the latent tail component of the model, $K_i$ is used as the
covariate governing the event-specific GPD scale. We use the same
formulation as \citet{Murphy2026spatio}, where there is strong evidence
that inclusion of this covariate improves the fit even when measurement error is ignored and that it results in a high-quality of model fit: a linear specification
\citep{eastoe2009modelling},
\begin{equation}
\sigma_{u,i}
=
g_L(K_i;\boldsymbol{\theta}_L)
=
\beta_0+\beta_1K_i, \qquad \text{for}~i=1,\ldots, n,
\label{sigmaGPD}
\end{equation}
where $\beta_0 \in \mathbb{R}$ controls the baseline GPD scale and $\beta_1 \in \mathbb{R}_{+}$
controls how the upper-tail scale changes with respect to Kaiser stress, with constraints
imposed to ensure $\sigma_{u+\delta,i}>0$ for all $i$. For implementation, the continuous range of $K_i$ values is divided
into six equal-width stress bins, following the idea described in
Section~\ref{subsec:covariate_discretisation}. Each event is assigned to a
stress bin according to its value of $K_i$, regardless of whether its
recorded magnitude lies above or below $u+\delta$. Within each stress
bin, the median of the $K_i$ values in that bin is used as the
representative stress value to define the corresponding latent density for magnitude in that covariate bin.

\subsubsection{Measurement error covariate: based on geophone network}
\label{subsubsec:groningen_geophones}

The uncertainty associated with a recorded magnitude is expected to
depend on how well the event is observed by the monitoring network. Figure~\ref{fig:groningen_summary_geo} illustrates that the Groningen geophone network has changed substantially over the catalogue
period, with the number and spatial coverage of active geophones
increasing over time \citep{KNMI1993}. The left panel shows a clear step change in the number of active geophones around 2016, corresponding to the expansion of the monitoring network. The two right panels show that the spatial coverage of the gas field has also improved, with the 2020 network providing more uniform coverage than the 2010 network. This motivates a measurement error model in which the noise
scale varies between events, rather than being constant over the full
catalogue.

In order to best summarise how measurement error varies over time, we exploit established geophysical theory to construct a suitable covariate, $N_{g,i}$, which is the number of active geophones within the detection radius of a given event. For event $i$, let $R_i$ denote the magnitude-dependent detection
radius, where $R_i$ can be computed from a source--path amplitude model for the predicted
P-wave signal and a prescribed background noise level \citep{kjartansson1979constant}. The details of
the calculation of $R_i$ are given in the supplementary material, in brief it captures several physical features that affect magnitude uncertainty, including the event magnitude, source--receiver distance, attenuation and wave propagation. Let $d_{ij}$
denote the distance between event $i$ and geophone $j$. Then $N_{g,i}$ is defined as
\[
N_{g,i}
=
\sum_j
\mathbbm{1}\{d_{ij}\le R_i\},
\]
where the sum is over the geophones that are active at the time of event $i$. 

We model the measurement error scale as a decreasing
function of $N_{g,i}$,
\begin{equation}
\sigma_{N,i}
=
g_N(N_{g,i};\boldsymbol{\theta}_N)
=
\frac{\sigma_{N_0}}{(N_{g,i}+1)^\gamma},
\qquad
\boldsymbol{\theta}_N=(\sigma_{N_0},\gamma),
\quad
\sigma_{N_0}>0,
\quad
0\leq\gamma\leq 1/2 .
\label{eq:noisescale}
\end{equation}
Here, $\sigma_{N_0}$ represents the maximum measurement error scale
under this specification, corresponding to the case $N_{g,i}=0$, and
$\gamma$ controls the rate at which the measurement error scale
decreases as the number of detecting geophones increases. The addition
of one to $N_{g,i}$ in expression~\eqref{eq:noisescale} ensures that $\sigma_{N,i}$ remains well-defined when $N_{g,i}=0$. We restrict $\gamma$ to the interval $0\leq\gamma\leq 1/2$ for the following reasons. The lower bound $\gamma=0$
corresponds to a constant measurement error scale, while the upper bound
$\gamma=1/2$ corresponds to the square-root reduction expected when
different geophone recordings contribute additively to the precision of
the magnitude estimate under an assumption of independent measurements
over the geophones. Estimating $\gamma$ within this interval allows the
data to determine the strength of the geophone count effect.

Figure~\ref{fig:groningen_covariates} (right) shows the scatter plot of $N_{g,i}$ against Kaiser stress
$K_i$. The regression smoother shown in the figure indicates that the two covariates are not independent. Their
dependence is partly driven by the temporal evolution of the catalogue: the number of detecting geophones increases
as the monitoring network expands, while Kaiser stress also increases over the catalogue period due to gas extraction.
The empirical Kendall correlation between $N_{g,i}$ and $K_i$ is
approximately $0.32$, indicating positive association. As shown in simulation scenario 3 in the supplementary material, dependence between the latent scale and measurement error covariates can make the two effects harder to distinguish and can affect inference for the model parameters. We therefore include regularisation on the measurement error component in the Groningen analysis in Section~\ref{subsection:inferenceres}. The penalty is used to represent external information about plausible magnitude uncertainty discussed in the Introduction and helps to stabilise estimation of the measurement error component. Since $\sigma_{N_0}$ is a positive scale parameter, we use a Gamma penalty with mean $0.3$ and standard deviation $0.15$. The mean is set at $0.3$ because this corresponds to the upper end of the magnitude uncertainty regarded as plausible by
the geophysical experts, which is appropriate here because
$\sigma_{N_0}$ represents the largest measurement error scale under the
model. The corresponding penalty density is shown in the supplementary material.

\subsection{Inference results and comparisons}
\label{subsection:inferenceres}
\subsubsection{Introduction}
To assess the effect of accounting for measurement error, this section presents results from two approaches: the proposed GPD-ME model and the na\"ive stress-dependent GPD model (subsequently termed stress-GPD) fitted to the recorded magnitudes while ignoring measurement error. Both models use the same Kaiser stress structure for the GPD scale as in expression~\eqref{sigmaGPD}, but only the GPD-ME models explicitly accounts for measurement uncertainty in the recorded magnitudes. In the GPD-ME model, the measurement error scale is additionally modelled as a function of the number of detecting geophones, $N_{g,i}$, as in expression~\eqref{eq:noisescale}. Inference uncertainty is assessed using a nonparametric bootstrap with $B=200$ bootstrap samples, each containing $1565$ events. For each event $i$, we resample the full event record
$(y_i,t_i,\boldsymbol{s}_i,K_i,N_{g,i}),
$
to preserve the empirical dependence between $Y_i$, $K_i$ and
$N_{g,i}$ within the catalogue. The two approaches are then refitted to each bootstrap catalogue using the same thresholds. The results are organised as follows. We first present the fitted GPD-ME model, including the parameter estimates, bootstrap-based confidence intervals and fitted covariate
effects. We then compare the GPD-ME model with the na\"ive stress-GPD model in terms of tail parameter estimates, fitted GPD scale, QQ-plots and return level estimates.

\subsubsection{Fitted GPD-ME model}
\label{subsubsec:app_gpd_me_fit}

As discussed in Section~\ref{subsubsec:groningen_geophones}, the positive
association between $N_{g,i}$ and $K_i$ can make the latent tail
scale and measurement error scale difficult to separate when fitting the GPD-ME. We therefore
fit the GPD-ME model using the regularised likelihood described in
Section~\ref{subsec:penalisation}, with the penalty applied to the baseline
measurement error scale $\sigma_{N_0}$.

\begin{table}[H]
\centering
\caption{
Point estimates and 95\% bootstrap-based confidence intervals for the
GPD-ME and stress-GPD model parameters in the Groningen application.
}
\label{tab:gron_gpdm_e_param_summary}

\setlength{\tabcolsep}{9pt}
\renewcommand{\arraystretch}{1.15}

\begin{tabular}{@{}lcccc@{}}
\hline
&
\multicolumn{2}{c}{GPD-ME} &
\multicolumn{2}{c@{}}{Stress-GPD} \\
\cline{2-3}
\cline{4-5}
Parameter &
Point estimate &
95\% CI &
Point estimate &
95\% CI \\
\hline
$\hat{\xi}$
& $-0.083$
& $(-0.140,\,-0.012)$
& $-0.101$
& $(-0.172,\,-0.055)$ \\

$\hat{\beta}_0$
& 0.254
& $(0.161,\,0.407)$
& -- 
& -- \\

$\hat{\beta}_1$
& 0.993
& $(0.461,\,1.426)$
& 0.742
& $(0.388,\,1.193)$ \\

$\hat{\sigma}_{N,0}$
& 0.197
& $(0.173,\,0.225)$
& --
& -- \\

$\hat{\gamma}$
& 0.360
& $(0.292,\,0.381)$
& --
& -- \\
\hline
\end{tabular}

\end{table}

Table~\ref{tab:gron_gpdm_e_param_summary} reports the point estimates
from the catalogue and the corresponding bootstrap-based 95\% confidence intervals. The fitted shape parameter is negative, with point estimate $\hat{\xi}=-0.083$ and with a confidence interval that excludes 0, supporting a finite upper endpoint for the latent
magnitude distribution. The estimate gives the fitted baseline
measurement error scale for an
event with no detecting geophones within the magnitude-dependent radius. This is within the range
$0.1$--$0.3\,M_L$ suggested by expert judgement for plausible
magnitude uncertainty at Groningen. The 95\% confidence interval for $\hat{\gamma}$ indicates
that the measurement error scale decreases with increasing geophone
support, with a decay rate considerably away from the constant-error case
$\gamma=0$ and the square-root rate $\gamma=1/2$. The inference for $\beta_1$ indicates that there is a statistical significant increase with Kaiser stress in the fitted GPD scale parameter above $u+\delta$.

\begin{figure}[H]
    \centering
    \includegraphics[width=0.89\linewidth]{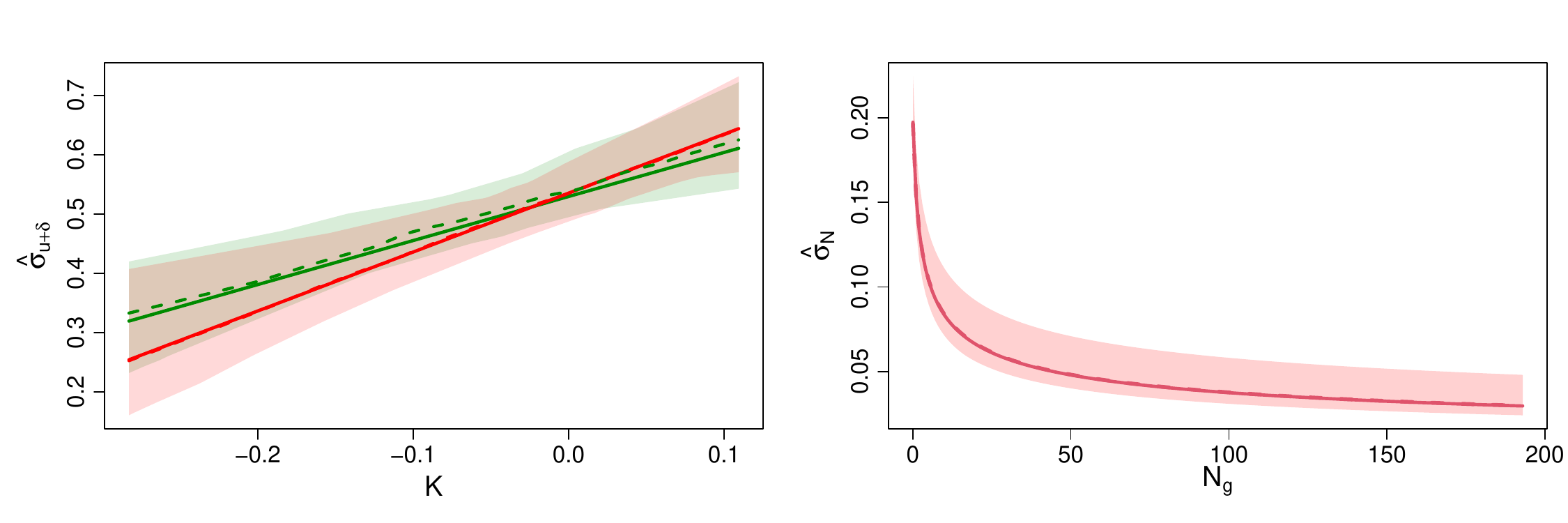}
\caption{
Fitted covariate effects and uncertainty in the Groningen application.
Left: fitted GPD scale $\sigma_{u+\delta}(K)$, comparing the
GPD-ME (red) and stress-GPD (green) models. Right: fitted measurement error scale $\sigma_N(N_g)$ under the GPD-ME
model. Solid lines show the original fits, dashed
lines show bootstrap medians, and shaded bands show pointwise
bootstrap-based 95\% confidence intervals.}
\label{fig:gpdme_covariate_bootstrap}
\end{figure}

Figure~\ref{fig:gpdme_covariate_bootstrap} shows the fitted covariate
effects across their domain. The left panel shows that the
fitted GPD scale above $u+\delta$ increases with $K$, indicating that larger Kaiser
stress values are associated with greater variability in latent threshold
exceedances. The bootstrap band summarises uncertainty in the fitted
stress-scale relationship. The right panel shows that even after accounting for inference uncertainty, the fitted measurement error scale $\sigma_N$ decreases from approximately
$0.20\,M_L$, when $N_g=0$, to around $0.05\,M_L$ for large values of
$N_g$. The reduction is steepest for small geophone counts and becomes
much more gradual once the number of detecting geophones is large. The estimated range overlaps substantially with the scientifically plausible
range of $0.10\,M_L$ to $0.30\,M_L$.

\subsubsection{Comparison with na\"ive GPD fitting}
\label{subsubsec:gpdme_comparison}

For the catalogue and each of the bootstrap samples described above, we also fit the (na\"ive) stress-GPD model to the recorded magnitudes $Y_i$,  with inference given in Table~\ref{tab:gron_gpdm_e_param_summary}. The GPD-ME and stress-GPD confidence intervals for $\xi$ have similar widths and partially overlap, but the GPD-ME interval is shifted towards less negative values of $\xi$. This suggests that ignoring measurement error leads to lower shape parameter inference, consistent with the simulation study in Section~\ref{Sec:Simulationstudy}. This difference should be interpreted together with changes in the fitted
GPD scale function $\sigma_{u+\delta}(K)$ as they are correlated \citep{chavez2005generalized}. As the shape and scale jointly determine the fitted tail probabilities above $u+\delta$, differences in both components can lead to different high-quantile,
return level and endpoint estimates, see Section~\ref{subsubsec:return_level_restimates}.

Figure~\ref{fig:gpdme_covariate_bootstrap} (left) compares the fitted stress-dependent GPD scale under the two models. This comparison is also
reflected in Table~\ref{tab:gron_gpdm_e_param_summary}, where the
estimated stress effect $\hat{\beta}_1$ is larger under the GPD-ME
model than under the stress-GPD model. The two approaches give similar patterns, with both estimates increasing as Kaiser stress increases, and the pointwise confidence intervals overlap across much of the covariate range. The largest difference appears at lower stress levels, where the GPD-ME model gives a smaller fitted latent GPD scale. This may be related to the structure of the catalogue: events with lower Kaiser stress are associated with smaller magnitudes, which have smaller detectability radii and hence fewer detecting geophones. The larger measurement uncertainty for these events allows the GPD-ME model to attribute less of the observed variation to the latent tail scale in this lower part of the Kaiser stress range. At larger Kaiser stress values, the pattern is reversed: the GPD-ME
model gives a slightly larger fitted GPD scale than the stress-GPD model. Together with the less negative fitted shape parameter under the GPD-ME model, these two differences imply larger fitted high quantile and
return level estimates at high $K$, and hence larger extrapolated risk estimates in the upper part of the Kaiser stress range, which is the important range for future risks.

\begin{figure}[H]
    \centering
    \includegraphics[width=0.65\linewidth]{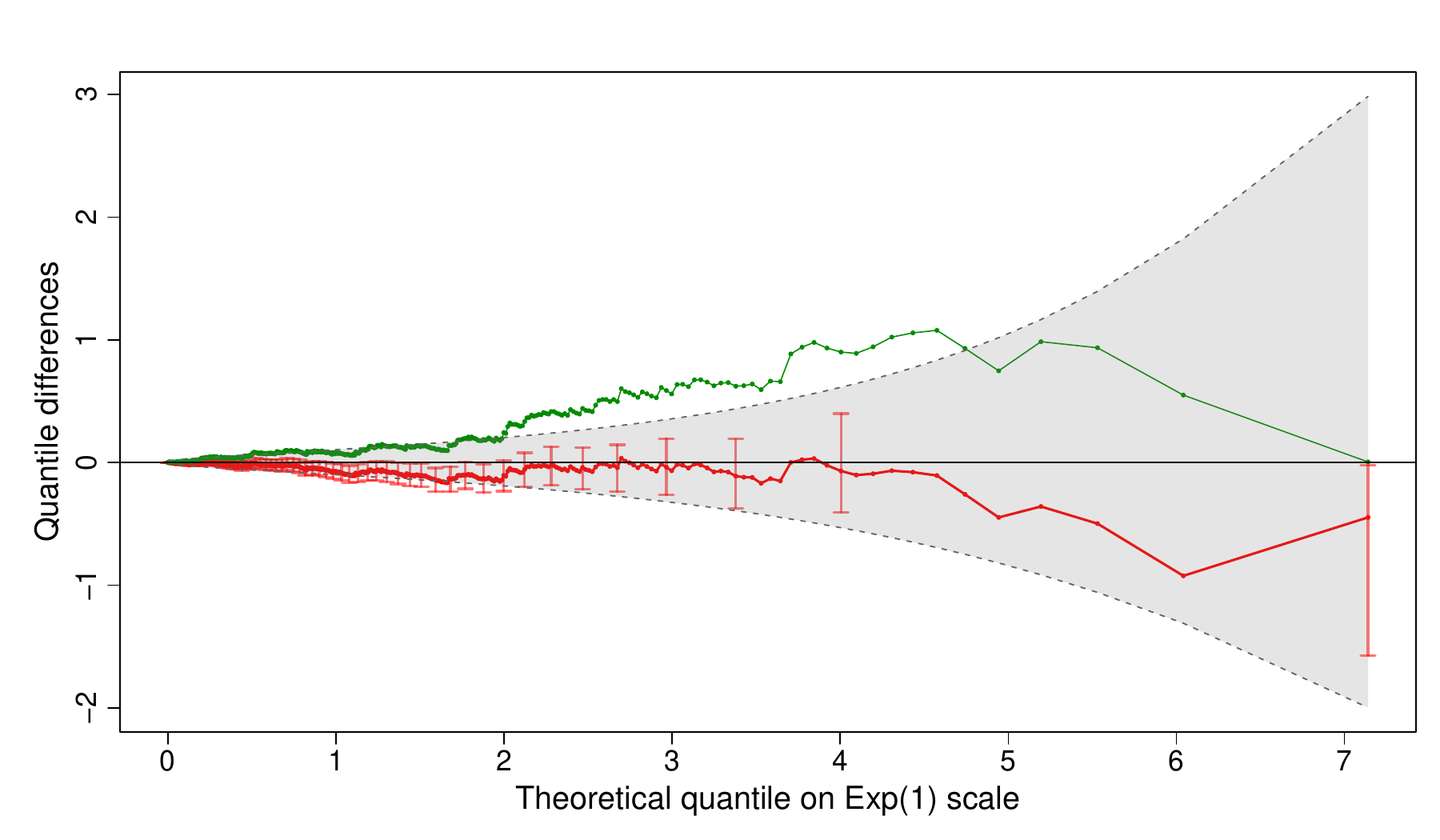}
    \caption{QQ difference plot on
the common $\mathrm{Exp}(1)$ scale. The GPD-ME curve (red) and the stress-GPD curve (green) show the QQ differences for the original fits. The grey region showing the pointwise
95\% tolerance intervals and vertical bars showing 95\% bootstrap-based confidence intervals for the GPD-ME fit.
}
    \label{fig:gpdme_comparison}
\end{figure}

Figure~\ref{fig:gpdme_comparison} compares the QQ diagnostics for the stress-GPD and GPD-ME fits using the exponential margin
QQ-difference scale introduced in Section~\ref{subsec: scenario1}. For each model, the observed exceedances $Y_i>u+\delta$ are transformed using the fitted conditional distribution above $u+\delta$, so that they should be identically and $\mathrm{Exp}(1)$ distributed if the fitted model was correct. The plotted quantity is the
difference between the ordered transformed values and the corresponding
$\mathrm{Exp}(1)$ quantiles. The stress-GPD fit, lies mostly above the upper tolerance intervals through much of the upper tail. This shows that the stress-GPD model underestimates the upper tail behaviour. In contrast, the GPD-ME curve is closer to zero and remains largely within the tolerance intervals. The bootstrap-based 95\% confidence intervals for GPD-ME also contain zero over most of the range, indicating that the GPD-ME model provides a good fit to the data.

\subsubsection{Return level estimates}
\label{subsubsec:return_level_restimates}
We finally examine how the differences in fitted GPD shape and scale between the two fits, affect
return-level estimates. Return levels are calculated from the fitted GPD
tail above $u+\delta$, using the stress-dependent scale function
$\sigma_{u+\delta}(K)$ and the fitted shape parameter from each model.
For the GPD-ME model, these return levels correspond to the fitted latent
magnitude distribution. Since the GPD scale depends on the latent covariate $Z_L$, the fitted conditional survival probability above a magnitude level $q>u+\delta$ is evaluated within each latent covariate bin. Let
\[
\bar F_r(q)
=
\Pr\{X>q \mid X>u+\delta,\; Z_L\in\mathcal{B}_{L,r}\},
\qquad r=1,\ldots,G_L,\quad q>u+\delta
\]
denote the fitted GPD conditional survival probability in bin $r$ for a selected distributional model for $X$. Using the stress-dependent scale function in expression~\eqref{sigmaGPD}, this is evaluated at the representative covariate value
$Z_{L,r}^{\mathrm{rep}}$, so that
\[
\bar F_r(q)
=
\left[
1+
\widehat{\xi}
\frac{q-(u+\delta)}
{g_L\!\left(Z_{L,r}^{\mathrm{rep}};
\widehat{\boldsymbol{\theta}}_L\right)}
\right]^{-1/\widehat{\xi}},
\]
with the usual
exponential limit when $\widehat{\xi}=0$. To allow the exceedance rate to vary across stress bins, let
$m_{u+\delta}^{[r]}$ denote the average annual rate of exceedances above $u+\delta$
when $Z_L\in\mathcal{B}_{L,r}$. The fitted annual exceedance rate above
a level $q$ is therefore
\[
\Lambda(q)
=
\sum_{r=1}^{G_L}
m_{u+\delta}^{[r]}\bar F_r(q).
\]
The $T$-year return level of the latent magnitude process, denoted by $a_T$, is then defined such that
$
T\Lambda(a_T)
=
1.
$
Thus, the return-level is obtained by solving this
equation for $a_T$ at each value of $T$. This gives a single
stress-weighted return level for the fitted process, while allowing both
the exceedance frequency and the GPD scale to vary across stress bins.

\begin{figure}[H]
\centering
\includegraphics[width=0.65\textwidth]{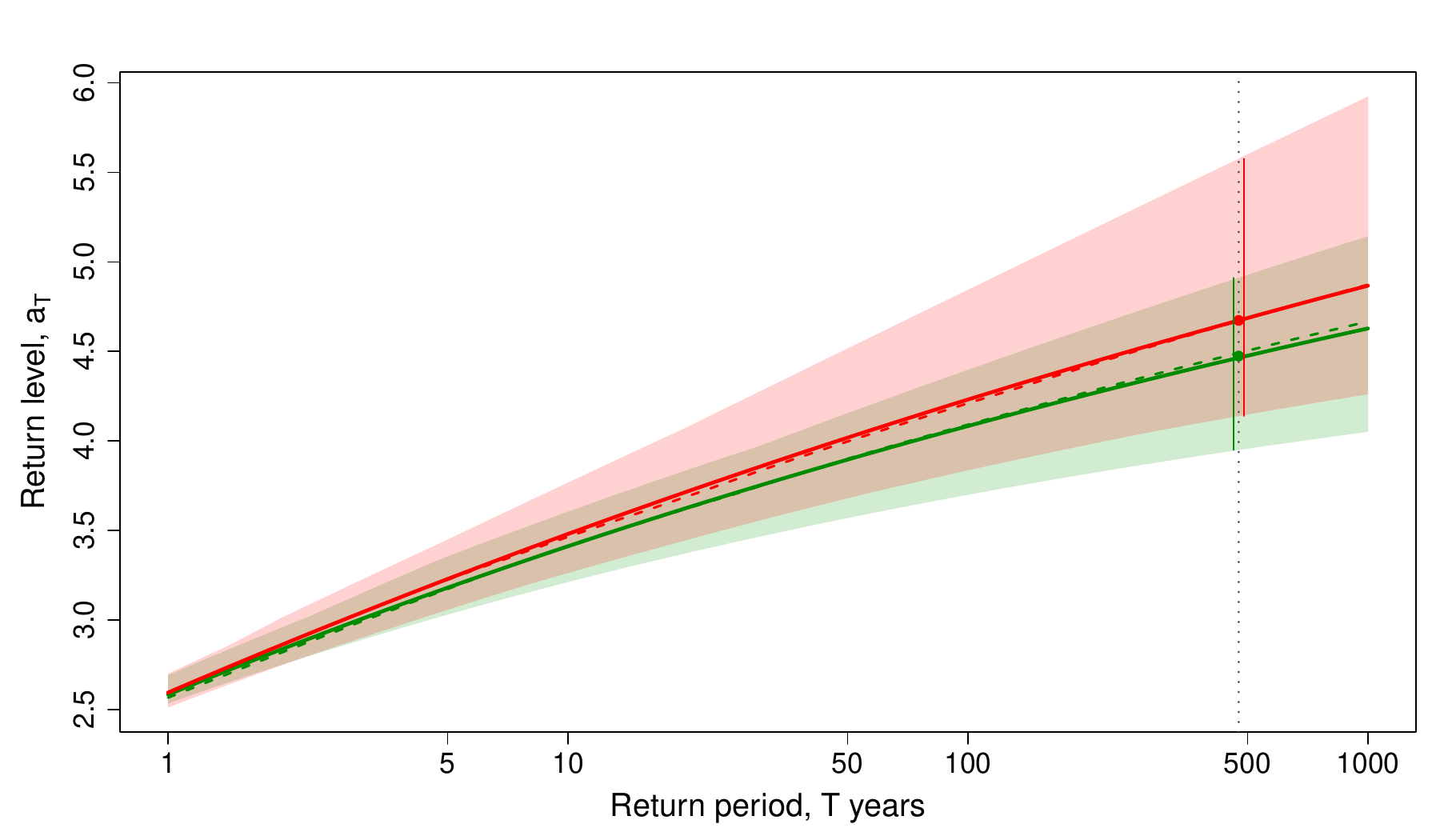}
\caption{
Comparison of stress-weighted return level curves $(T, a_T)$ from the stress-GPD model (green) and the GPD-ME model (red). Solid lines indicate the original fitted return levels, dashed lines indicate bootstrap medians, and shaded regions indicate bootstrap-based pointwise 95\% confidence intervals. The vertical dotted line marks the $475$-year return period.
}
\label{fig:return_level_stress_avg}
\end{figure}

Figure~\ref{fig:return_level_stress_avg} compares the resulting stress-weighted return level curves for the GPD-ME and na\"ive stress-GPD models. For both approaches, the estimated return level and the width of the bootstrap interval increase with the return period, reflecting the greater uncertainty associated with extrapolation to rarer events. The GPD-ME curve lies above the na\"ive stress-GPD curve and has wider confidence intervals for longer return periods, indicating that accounting for measurement error leads to larger estimated upper-tail quantiles and greater uncertainty in this application. These are important as it indicates that current practice, of ignoring measurement error is giving point estimates and uncertainties that exhibit underestimation of the level of risk for future earthquakes. This finding is at odds with the often held perspective that ignoring measurement error introduces conservatism into return level inference.

The vertical line marks the $475$-year return period. This return period is commonly used in seismic design because a level with a $10\%$ probability of exceedance over a $50$-year period corresponds approximately to a $475$-year return level under a stationary Poisson exceedance assumption \citep{code2005eurocode}. Using stress-dependent annual exceedance rates, the $475$-year return level estimate is $4.67\,M_{L}$ for the GPD-ME model, with a 95\% confidence interval $(4.14,5.58)$. The corresponding values for the stress-GPD model are $4.47\,M_{L}$ and $(3.95,4.91)$. The higher estimate and wider interval under the GPD-ME model are the natural reflection from the additional uncertainty introduced when inference is made for the latent magnitude distribution rather than incorrectly treating the recorded magnitudes as error-free, and as shown in Table~\ref{tab:case1_coverage_all} should substantially improve the coverage properties of associated confidence intervals.

\section{Discussion}
\label{Sec:discussion}
This paper develops a measurement error framework for upper-tail
inference when the recorded extreme observations are noisy versions of
latent quantities of scientific interest. The proposed approach, termed the GPD-ME, combines a hybrid
body--tail model for the latent distribution with an additive Gaussian
measurement error model, and uses a partially censored convolution
likelihood to focus inference on the upper tail. The framework also
allows both the latent GPD scale and the measurement error scale to vary
with covariates, which is important in applications where tail behaviour
and observation quality change across events. The simulation study shows
that ignoring measurement error (the na\"ive method) can bias inference for the GPD shape and
scale parameters, particularly as the signal-to-noise ratio decreases,
and can lead to poor coverage of confidence intervals for tail parameters
and high quantiles. In the settings considered here, the na\"ive method tends to give negatively biased
estimates of the GPD shape parameter and positively biased estimates of the GPD
scale, although the direction of this bias should not be assumed to be
universal. In the Groningen application, the GPD-ME gives estimates of the GPD shape parameter that are less negative than the na\"ive estimate and
produces a different fitted stress-dependent scale function, both features being supported by simulation studies. These differences lead to larger return-level estimates, while the estimated noise variability is consistent with geophysical expectations. The GPD-ME fit also gives improved QQ-plot
diagnostics, indicating better agreement between the fitted latent tail
model and the observed upper-tail behaviour after allowing for
measurement error.

Several methodological choices are important for making the proposed
framework practical. The convolution formulation directly represents the
relationship between the latent variable $X_i$ and the noisy observation
$Y_i$, so that measurement error is propagated into tail parameter and
high-quantile inference. The partially
censored likelihood further focuses inference on the upper tail by
allowing observations below the threshold to contribute only through their
probability of falling below the threshold, rather than through their
individual density values. Since
direct evaluation of the convolution terms in this likelihood can be
computationally expensive, we use FFT-based numerical evaluation, which
makes repeated likelihood evaluation feasible. Finally,
when the covariates controlling the latent GPD scale and the measurement
error scale are correlated, regularisation of the measurement error
component can help stabilise estimation and reduce confounding between
the two scale effects.

Robust methods for extreme value inference provide a related but distinct
approach. For example, \citet{dupuis1998robust} and
\citet{dupuis2006robust} reduce the effect of outlying observations
when estimating extreme value tails. 
Such methods are valuable when extreme observations may be contaminated
or when a small number of points have excessive influence on the fitted
tail. The goal here is different: the recorded observations are not
treated as outliers to be downweighted, but as noisy measurements of
latent values. Our proposed method explicitly estimates the measurement
error component and carries its effect through to uncertainty in the tail-parameter and
high-quantile estimates. A useful future comparison would be to assess
when robust tail estimators and explicit measurement error models give
similar protection, and when estimating the error process gives additional
benefits.

There are also connections with work on forecast verification and scoring
in the presence of observation error. Methods such as those of
\citet{ferro2017measuring} and \citet{bessac2021forecast} adjust forecast
evaluation when the verifying observations are themselves uncertain. The
common theme is that the recorded value should not be treated as the
truth when the observation process is uncertain. However, the aim of
the present paper is different: rather than correcting a forecast score,
we estimate the latent extreme value distribution from noisy observations. Thus, the measurement error model enters the
likelihood for tail inference, not only the evaluation criterion.

Several limitations and extensions remain. The most important concerns
the treatment of the threshold. In this paper, two thresholds are used to cover the continuous density transition from the body to the GPD tail of the hybrid distribution. The lower and upper thresholds are fixed before applying
the measurement error model. In many applications, however, the effective
threshold may vary over time or across space, e.g., in the induced earthquake context, because of
changes in network sensitivity, reporting practice or detection
capability \citep{Murphy2026spatio}. If such a threshold is known, the likelihood could be modified by replacing the fixed upper threshold $u+\delta$ with an
observation-specific threshold $v_i=u_i+\delta_i$. The partially censored likelihood would become: 
\[
L(\boldsymbol{\theta})
=
\prod_{i:y_i \le v_i}
F_{Y_i}
\left(
v_i;\boldsymbol{\theta},\boldsymbol{Z}_i
\right)
\prod_{i:y_i > v_i}
f_{Y_i}
\left(
y_i;\boldsymbol{\theta},\boldsymbol{Z}_i
\right).
\]
The convolution density $f_{Y_i}$ would then be evaluated using a
latent hybrid distribution whose GPD tail begins at $v_i$. If a common
shape parameter is assumed, the GPD scale at $v_i$ could be linked to a
reference threshold $u$ through the threshold stability property, $\sigma_{v_i}=\sigma_{u}+\xi\{v_i-u\}$, for $v_i>u$. This extension would also create additional computational cost, since
the convolution may need to be evaluated at many different threshold and
covariate values. A binning strategy similar to that described in
Section~\ref{subsec:covariate_discretisation} could be used to make the
likelihood evaluation feasible.

The role of the body distribution is more delicate. If the threshold
varies only because of the observation process, but the latent
distribution is otherwise assumed to be identically distributed over $i$, then a common latent
body distribution could still be used, with different observations treated as being censored in the likelihood at different values. If the observed
data also exhibits evidence of being non-identically distributed, the body distribution
may need to vary over time or space as well. Therefore,  extending the method to this setting would require a joint treatment of
the latent value, the varying threshold and the GPD parameters. The main
difficulty is identifiability. Variation in the observed upper tail could
be explained by changes in the threshold, changes in the latent GPD
scale, changes in the measurement error scale, or changes in the body
distribution near the threshold. 

More problematic is when the threshold itself is uncertain. One
possible approach is to select a threshold function before applying the
GPD-ME method, e.g., using existing threshold-selection methods for
non-identically distributed extreme value models \citep{varty2021inference, Murphy2026spatio}. However, when these methods are
applied to the noisy observations $Y_i$, the selected threshold
is the best value for the observed distribution rather than for the latent distribution of interest. A more comprehensive treatment would model the threshold jointly with the
tail and measurement error parameters, or assign a distribution to the
threshold and integrate over this uncertainty. 

A further computational extension concerns the binning used in the FFT likelihood evaluation. In this paper, the covariate likelihood is approximated by dividing the covariate space into bins, estimating the KDE body component within
latent covariate bins, and evaluating the convolution at representative covariate values. This reduces the computational cost because the convolution need only be evaluated once per each bin. However, the binning approximation
can introduce bias when the bin-specific KDE body component or the scale
functions vary substantially within a bin. A possible improvement would be to evaluate the convolution on a finer grid before fitting the model, and then use interpolation or a smooth Gaussian process emulator to approximate the likelihood during fitting.

Other extensions concern the measurement error model itself. The Gaussian error assumption is convenient and provides a simple first
approximation to the combined effect of multiple sources of magnitude
uncertainty as well as being broadly supported by geophysicists for the Groningen data. However, real observation processes may involve features not captured by the present Gaussian error model, such as skewed or heavy-tailed
measurement errors, or rounding of the recorded observations. The same
convolution-based framework could be used with alternative error
distributions. A second extension concerns settings with very low signal-to-noise ratios, where the measurement error scale is large relative to the latent tail scale. In this case, the convolution can produce very similar
observed tail densities from different combinations of latent GPD scale and noise scale. The likelihood may then contain a weakly identified subspace with high but flat likelihood, making it difficult to separate the latent tail
scale from the measurement error scale. External information about the
error scale, stronger regularisation or repeated measurements may then be
needed to obtain stable tail inference or possibly the use of PP fits, like in Figure~\ref{fig:sim_case1_naive_qq}, to identify if the GPD-ME is suitable or needed.

Finally, the proposed framework could be extended beyond threshold
exceedances. For block maxima, a related approach would model the latent
block maxima using a GEV distribution and convolve this distribution with
an measurement error model. This would provide an analogue of the GPD-ME
method for block maxima. However, additional care
would be needed when the observed maximum may not correspond to the same
event as the latent maximum after measurement error is added. This makes the block maxima setting conceptually different from modelling
noisy threshold exceedances: measurement error may affect not only the
recorded value of the block maximum, but also which event is selected as
the maximum.

Overall, the results demonstrate that measurement error can have a strong effect on extreme value inference, even when the error scale is moderate relative to the latent variability. Accounting for this error improves the calibration of uncertainty intervals and gives tail inferences that are more closely aligned with the latent distribution of
interest. For the Groningen catalogue, the method provides a way to link changes in the monitoring network and geophysical covariates to uncertainty in extreme magnitude estimates. More generally, the approach is relevant for environmental extremes and other applications where observations are noisy, detection capability varies over time, and risk assessment depends on extrapolation beyond the range of the observed data.

\section*{Declaration of Competing Interest}
The authors declare that they have no known competing financial interests or personal relationships that could have appeared to influence the work reported in this paper.

\section*{Acknowledgements}
 The authors also would like to thank Stephen Bourne and Steve Oates (Shell Research Ltd) for providing the support and supplying the earthquake catalogue. This paper is based on work completed while Wanchen Yue was part of the EPSRC funded STOR-i centre for doctoral training (EP/S022252/1).

\bibliographystyle{agsm}
\bibliography{ref}
\end{document}